\documentclass[preprint,twocolumn]{aastex7}

\begin{document}
\title{HAWC Study on the Ultra-High-Energy Gamma-Ray Emissions from the Pulsar Wind Nebula G32.64+0.53}

\correspondingauthor{R.~Babu, C.D.~Rho, Y.~Son}

\author{R.~Alfaro}
\affiliation{{Instituto de F\'{i}sica, Universidad Nacional Autónoma de México, Ciudad de Mexico, Mexico }}
\email{ruben@fisica.unam.mx}

\author[0000-0001-8310-4486]{C.~Alvarez}
\affiliation{{Universidad Autónoma de Chiapas, Tuxtla Gutiérrez, Chiapas, México}}
\email{crabpulsar@hotmail.com}

\author{E.~Anita-Rangel}
\affiliation{{Instituto de Astronom\'{i}a, Universidad Nacional Autónoma de México, Ciudad de Mexico, Mexico }}
\email{earangel@astro.unam.mx}

\author[0000-0002-0595-9267]{M.~Araya}
\affiliation{{Universidad de Costa Rica, San José 2060, Costa Rica}}
\email{miguel.araya@ucr.ac.cr}

\author{J.C.~Arteaga-Velázquez}
\affiliation{{Universidad Michoacana de San Nicolás de Hidalgo, Morelia, Mexico }}
\email{juan.arteaga@umich.mx}

\author[0000-0002-4020-4142]{D.~Avila Rojas}
\affiliation{{Instituto de Astronom\'{i}a, Universidad Nacional Autónoma de México, Ciudad de Mexico, Mexico }}
\email{doavila@astro.unam.mx}

\author[0000-0002-2084-5049]{H.A.~Ayala Solares}
\affiliation{{Temple University, Department of Physics, 1925 N. 12th Street, Philadelphia, PA 19122, USA}}
\email{hugo.ayala.solares@temple.edu}

\author[0000-0002-5529-6780]{R.~Babu}
\affiliation{Department of Physics and Astronomy, Michigan State University, East Lansing, MI, USA} 
\email[show]{baburish@msu.edu}

\author[0000-0002-3886-3739]{P.~Bangale}
\affiliation{{Temple University, Department of Physics, 1925 N. 12th Street, Philadelphia, PA 19122, USA}}
\email{priyadarshini.bangale@temple.edu}

\author[0000-0003-3207-105X]{E.~Belmont-Moreno}
\affiliation{{Instituto de F\'{i}sica, Universidad Nacional Autónoma de México, Ciudad de Mexico, Mexico }}
\email{belmont@fisica.unam.mx}

\author[0000-0002-6781-4004]{A.~Bernal}
\affiliation{{Instituto de Astronom\'{i}a, Universidad Nacional Autónoma de México, Ciudad de Mexico, Mexico }}
\email{abel@astro.unam.mx}

\author[0000-0002-4042-3855]{K.S.~Caballero-Mora}
\affiliation{{Universidad Autónoma de Chiapas, Tuxtla Gutiérrez, Chiapas, México}}
\email{karen.scm@gmail.com}

\author[0000-0003-2158-2292]{T.~Capistrán}
\affiliation{Università degli Studi di Torino, I-10125 Torino, Italy}
\email{tcapistranc@gmail.com}

\author[0000-0002-8553-3302]{A.~Carramiñana}
\affiliation{{Instituto Nacional de Astrof\'{i}sica, Óptica y Electrónica, Puebla, Mexico }}
\email{alberto@inaoep.mx}

\author{F.~Carreón}
\affiliation{{Instituto de Astronom\'{i}a, Universidad Nacional Autónoma de México, Ciudad de Mexico, Mexico }}
\email{mfcarreon@astro.unam.mx}

\author[0000-0002-6144-9122]{S.~Casanova}
\affiliation{{Institute of Nuclear Physics Polish Academy of Sciences, PL-31342 IFJ-PAN, Krakow, Poland }}
\email{sabrinacasanova@gmail.com}

\author{U.~Cotti}
\affiliation{{Universidad Michoacana de San Nicolás de Hidalgo, Morelia, Mexico }}
\email{umberto.cotti@umich.mx}

\author[0000-0002-1132-871X]{J.~Cotzomi}
\affiliation{{Facultad de Ciencias F\'{i}sico Matemáticas, Benemérita Universidad Autónoma de Puebla, Puebla, Mexico }}
\email{jcotzomi@yahoo.com.mx}

\author[0000-0002-7747-754X]{S.~Coutiño de León}
\affiliation{{Instituto de Física Corpuscular, CSIC, Universitat de València, E-46980, Paterna, Valencia, Spain}}
\email{sara.cdl989@gmail.com}

\author[0000-0001-9643-4134]{E.~De la Fuente}
\affiliation{{Departamento de F\'{i}sica, Centro Universitario de Ciencias Exactase Ingenierias, Universidad de Guadalajara, Guadalajara, Mexico }}
\email{edfuente@gmail.com}

\author[0000-0001-9768-1858]{P.~Desiati}
\affiliation{{Dept. of Physics and Wisconsin IceCube Particle Astrophysics Center, University of Wisconsin{\textemdash}Madison, Madison, WI, USA}}
\email{paolo.desiati@icecube.wisc.edu}

\author{N.~Di Lalla}
\affiliation{{Department of Physics, Stanford University: Stanford, CA 94305–4060, USA}}
\email{niccolo.dilalla@stanford.edu}

\author[0000-0001-8487-0836]{R.~Diaz Hernandez}
\affiliation{{Instituto Nacional de Astrof\'{i}sica, Óptica y Electrónica, Puebla, Mexico }}
\email{dihera77@gmail.com}

\author[0000-0002-2987-9691]{M.A.~DuVernois}
\affiliation{{Dept. of Physics and Wisconsin IceCube Particle Astrophysics Center, University of Wisconsin{\textemdash}Madison, Madison, WI, USA}}
\email{duvernois@icecube.wisc.edu}

\author[0000-0002-0087-0693]{J.C.~Díaz-Vélez}
\affiliation{{Dept. of Physics and Wisconsin IceCube Particle Astrophysics Center, University of Wisconsin{\textemdash}Madison, Madison, WI, USA}}
\email{juancarlos@icecube.wisc.edu}

\author[0000-0001-5737-1820]{K.~Engel}
\affiliation{{Department of Physics, University of Maryland, College Park, MD, USA }}
\email{kristi.engel23@gmail.com}

\author[0000-0003-2423-4656]{T.~Ergin}
\affiliation{{Department of Physics and Astronomy, Michigan State University, East Lansing, MI, USA }}
\email{ergin.tulun@gmail.com}

\author[0000-0001-7074-1726]{C.~Espinoza}
\affiliation{{Instituto de F\'{i}sica, Universidad Nacional Autónoma de México, Ciudad de Mexico, Mexico }}
\email{m.catalina@fisica.unam.mx}

\author[0000-0002-0173-6453]{N.~Fraija}
\affiliation{{Instituto de Astronom\'{i}a, Universidad Nacional Autónoma de México, Ciudad de Mexico, Mexico }}
\email{nifraija@astro.unam.mx}

\author{S.~Fraija}
\affiliation{{Instituto de Astronom\'{i}a, Universidad Nacional Autónoma de México, Ciudad de Mexico, Mexico }}
\email{sarafraija@hotmail.com}

\author[0000-0002-4188-5584]{J.A.~García-González}
\affiliation{{Tecnologico de Monterrey, Escuela de Ingenier\'{i}a y Ciencias, Ave. Eugenio Garza Sada 2501, Monterrey, N.L., Mexico, 64849}}
\email{anteus79@tec.mx}

\author{F.~Garfias}
\affiliation{{Instituto de Astronom\'{i}a, Universidad Nacional Autónoma de México, Ciudad de Mexico, Mexico }}
\email{fergar@astro.unam.mx}

\author{N.~Ghosh}
\affiliation{{Department of Physics, Michigan Technological University, Houghton, MI, USA }}
\email{nghosh1@mtu.edu}

\author[0000-0002-5209-5641]{M.M.~González}
\affiliation{{Instituto de Astronom\'{i}a, Universidad Nacional Autónoma de México, Ciudad de Mexico, Mexico }}
\email{magda@astro.unam.mx}

\author{J.A.~González}
\affiliation{{Universidad Michoacana de San Nicolás de Hidalgo, Morelia, Mexico }}
\email{jose.gonzalez.c@umich.mx}

\author[0000-0002-9790-1299]{J.A.~Goodman}
\affiliation{{Department of Physics, University of Maryland, College Park, MD, USA }}
\email{goodman@umd.edu}

\author[0000-0002-0870-2328]{D.~Guevel}
\affiliation{{Department of Physics, Michigan Technological University, Houghton, MI, USA }}
\email{djguevel@mtu.edu}

\author[0009-0003-8844-6321]{J.~Gyeong}
\affiliation{{Department of Physics, Sungkyunkwan University, Suwon 16419, South Korea}}
\email{kyoungjh1011@naver.com}

\author[0000-0001-9844-2648]{J.P.~Harding}
\affiliation{{Los Alamos National Laboratory, Los Alamos, NM, USA }}
\email{jpharding@lanl.gov}

\author[0000-0001-5169-723X]{I.~Herzog}
\affiliation{{Department of Physics and Astronomy, Michigan State University, East Lansing, MI, USA }}
\email{herzogia@msu.edu}

\author[0000-0002-5447-1786]{D.~Huang}
\affiliation{{Department of Physics and Astronomy, University of Delaware, Newark, DE, USA}}
\email{dezhih@mtu.edu}

\author[0000-0002-5527-7141]{F.~Hueyotl-Zahuantitla}
\affiliation{{Universidad Autónoma de Chiapas, Tuxtla Gutiérrez, Chiapas, México}}
\email{filihz@gmail.com}

\author[0000-0002-3302-7897]{P.~H\"{u}ntemeyer}
\affiliation{Department of Physics, Michigan Technological University, Houghton, MI, USA}
\email{petra@mtu.edu}

\author[0000-0001-5811-5167]{A.~Iriarte}
\affiliation{{Instituto de Astronom\'{i}a, Universidad Nacional Autónoma de México, Ciudad de Mexico, Mexico }}
\email{airiarte@astro.unam.mx}

\author{S.~Kaufmann}
\affiliation{{Universidad Politecnica de Pachuca, Pachuca, Hgo, Mexico }}
\email{skaufmann13@googlemail.com}

\author[0000-0003-4785-0101]{D.~Kieda}
\affiliation{{Department of Physics and Astronomy, University of Utah, Salt Lake City, UT, USA }}
\email{dave.kieda@utah.edu}

\author[0009-0005-8773-6057]{K.~Leavitt}
\affiliation{{Department of Physics, Michigan Technological University, Houghton, MI, USA }}
\email{kleavitt@mtu.edu}

\author[0000-0002-2467-5673]{W.H.~Lee}
\affiliation{{Instituto de Astronom\'{i}a, Universidad Nacional Autónoma de México, Ciudad de Mexico, Mexico }}
\email{wlee@astro.unam.mx}

\author[0000-0002-2153-1519]{J.~Lee}
\affiliation{{University of Seoul, Seoul, Rep. of Korea}}
\email{jason.lee@uos.ac.kr}

\author[0000-0003-0513-3841]{C.~de León}
\affiliation{{Universidad Michoacana de San Nicolás de Hidalgo, Morelia, Mexico }}
\email{cederik.de.leon@umich.mx}

\author[0000-0001-5516-4975]{H.~León Vargas}
\affiliation{{Instituto de F\'{i}sica, Universidad Nacional Autónoma de México, Ciudad de Mexico, Mexico }}
\email{hleonvar@fisica.unam.mx}

\author[0000-0001-8825-3624]{A.L.~Longinotti}
\affiliation{{Instituto de Astronom\'{i}a, Universidad Nacional Autónoma de México, Ciudad de Mexico, Mexico }}
\email{alonginotti@astro.unam.mx}

\author[0000-0003-2810-4867]{G.~Luis-Raya}
\affiliation{{Universidad Politecnica de Pachuca, Pachuca, Hgo, Mexico }}
\email{gilura6969@hotmail.com}

\author[0000-0001-8088-400X]{K.~Malone}
\affiliation{{Los Alamos National Laboratory, Los Alamos, NM, USA }}
\email{kmalone@lanl.gov}

\author{O.~Martinez}
\affiliation{{Facultad de Ciencias F\'{i}sico Matemáticas, Benemérita Universidad Autónoma de Puebla, Puebla, Mexico }}
\email{omartin@fcfm.buap.mx}

\author[0000-0002-2824-3544]{J.~Martínez-Castro}
\affiliation{{Centro de Investigaci\'on en Computaci\'on, Instituto Polit\'ecnico Nacional, M\'exico City, M\'exico.}}
\email{macj@cic.ipn.mx}

\author[0000-0002-2610-863X]{J.A.~Matthews}
\affiliation{{Dept of Physics and Astronomy, University of New Mexico, Albuquerque, NM, USA }}
\email{johnm@unm.edu}

\author[0000-0002-8390-9011]{P.~Miranda-Romagnoli}
\affiliation{{Universidad Autónoma del Estado de Hidalgo, Pachuca, Mexico }}
\email{pa.miranda.r@gmail.com}

\author{J.A.~Morales-Soto}
\affiliation{{Universidad Michoacana de San Nicolás de Hidalgo, Morelia, Mexico }}
\email{jmoralessg@gmail.com}

\author[0000-0002-1114-2640]{E.~Moreno}
\affiliation{{Facultad de Ciencias F\'{i}sico Matemáticas, Benemérita Universidad Autónoma de Puebla, Puebla, Mexico }}
\email{emoreno@fcfm.buap.mx}

\author[0000-0002-7675-4656]{M.~Mostafá}
\affiliation{{Temple University, Department of Physics, 1925 N. 12th Street, Philadelphia, PA 19122, USA}}
\email{miguel@psu.edu}

\author{M.~Najafi}
\affiliation{{Department of Physics, Michigan Technological University, Houghton, MI, USA }}
\email{mnajafi@mtu.edu}

\author{A.~Nayerhoda}
\affiliation{{Institute of Nuclear Physics Polish Academy of Sciences, PL-31342 IFJ-PAN, Krakow, Poland }}
\email{amid.nayerhoda@gmail.com}

\author[0000-0003-1059-8731]{L.~Nellen}
\affiliation{{Instituto de Ciencias Nucleares, Universidad Nacional Autónoma de Mexico, Ciudad de Mexico, Mexico }}
\email{lukas@nucleares.unam.mx}

\author[0000-0001-7099-108X]{R.~Noriega-Papaqui}
\affiliation{{Universidad Autónoma del Estado de Hidalgo, Pachuca, Mexico }}
\email{ropapaqui@gmail.com}

\author[0000-0002-5448-7577]{N.~Omodei}
\affiliation{{Department of Physics, Stanford University: Stanford, CA 94305–4060, USA}}
\email{nicola.omodei@stanford.edu}

\author[0000-1111-2222-3333]{E.~Ponce}
\affiliation{{Facultad de Ciencias F\'{i}sico Matemáticas, Benemérita Universidad Autónoma de Puebla, Puebla, Mexico }}
\email{eponce@fcfm.buap.mx}

\author[0000-0002-8774-8147]{Y.~Pérez Araujo}
\affiliation{{Instituto de F\'{i}sica, Universidad Nacional Autónoma de México, Ciudad de Mexico, Mexico }}
\email{yuniorpy@gmail.com}

\author[0000-0001-5998-4938]{E.G.~Pérez-Pérez}
\affiliation{{Universidad Politecnica de Pachuca, Pachuca, Hgo, Mexico }}
\email{egperezp@yahoo.com.mx}

\author[0000-0002-8940-5316]{A.~Pratts}
\affiliation{{Instituto de F\'{i}sica, Universidad Nacional Autónoma de México, Ciudad de Mexico, Mexico}}
\email{yoba_m_t_a@ciencias.unam.mx}

\author[0000-0002-1858-2622]{S.~Recchia}
\affiliation{{Institute of Nuclear Physics Polish Academy of Sciences, PL-31342 IFJ-PAN, Krakow, Poland }}
\email{sarah.recchia@ifj.edu.pl}

\author[0000-0002-6524-9769]{C.D.~Rho}
\affiliation{Department of Physics, Sungkyunkwan University, Suwon 16419, Republic of Korea} 
\email[show]{cdr397@skku.edu}

\author{A.~Rodriguez Parra}
\affiliation{{Universidad Michoacana de San Nicolás de Hidalgo, Morelia, Mexico }}
\email{ancelmo.rodriguez@umich.mx}

\author[0000-0003-1327-0838]{D.~Rosa-González}
\affiliation{{Instituto Nacional de Astrof\'{i}sica, Óptica y Electrónica, Puebla, Mexico }}
\email{danrosa@inaoep.mx}

\author{M.~Roth}
\affiliation{{Los Alamos National Laboratory, Los Alamos, NM, USA }}
\email{mattroth@lanl.gov}

\author[0000-0003-4556-7302]{H.~Salazar}
\affiliation{{Facultad de Ciencias F\'{i}sico Matemáticas, Benemérita Universidad Autónoma de Puebla, Puebla, Mexico }}
\email{hsalazar@fcfm.buap.mx}

\author[0000-0002-9312-9684]{D.~Salazar-Gallegos}
\affiliation{{Department of Physics and Astronomy, Michigan State University, East Lansing, MI, USA }}
\email{salaza82@msu.edu}

\author[0000-0001-6079-2722]{A.~Sandoval}
\affiliation{{Instituto de F\'{i}sica, Universidad Nacional Autónoma de México, Ciudad de Mexico, Mexico }}
\email{asandoval@fisica.unam.mx}

\author[0000-0001-8644-4734]{M.~Schneider}
\affiliation{{Department of Physics, University of Maryland, College Park, MD, USA }}
\email{mschnei4@umd.edu}

\author{J.~Serna-Franco}
\affiliation{{Instituto de F\'{i}sica, Universidad Nacional Autónoma de México, Ciudad de Mexico, Mexico }}
\email{j_serna@ciencias.unam.mx}

\author{M.~Shin}
\affiliation{{Department of Physics, Sungkyunkwan University, Suwon 16419, South Korea}}
\email{minjishin23@gmail.com}

\author[0000-0002-1012-0431]{A.J.~Smith}
\affiliation{{Department of Physics, University of Maryland, College Park, MD, USA }}
\email{asmith8@umd.edu}

\author[0000-0002-7214-8480]{Y.~Son}
\affiliation{University of Seoul, Seoul, Republic of Korea} 
\email[show]{youngwan.son@cern.ch}

\author[0000-0002-1492-0380]{R.W.~Springer}
\affiliation{{Department of Physics and Astronomy, University of Utah, Salt Lake City, UT, USA }}
\email{wayne.springer@utah.edu}

\author[0000-0002-9074-0584]{O.~Tibolla}
\affiliation{{Universidad Politecnica de Pachuca, Pachuca, Hgo, Mexico }}
\email{omar.tibolla@gmail.com}

\author[0000-0001-9725-1479]{K.~Tollefson}
\affiliation{{Department of Physics and Astronomy, Michigan State University, East Lansing, MI, USA }}
\email{tollefson@pa.msu.edu}

\author[0000-0002-1689-3945]{I.~Torres}
\affiliation{{Instituto Nacional de Astrof\'{i}sica, Óptica y Electrónica, Puebla, Mexico }}
\email{ibrahim.torres23@gmail.com}

\author[0000-0002-7102-3352]{R.~Torres-Escobedo}
\affiliation{{Tsung-Dao Lee Institute \& School of Physics and Astronomy, Shanghai Jiao Tong University, 800 Dongchuan Rd, Shanghai, SH 200240, China}}
\email{torresramiro350@sjtu.edu.cn}

\author[0000-0003-0715-7513]{E.~Varela}
\affiliation{{Facultad de Ciencias F\'{i}sico Matemáticas, Benemérita Universidad Autónoma de Puebla, Puebla, Mexico }}
\email{enrique.varela@correo.buap.mx}

\author[0000-0001-6876-2800]{L.~Villaseñor}
\affiliation{{Facultad de Ciencias F\'{i}sico Matemáticas, Benemérita Universidad Autónoma de Puebla, Puebla, Mexico }}
\email{lvillasen@gmail.com}

\author[0000-0001-6798-353X]{X.~Wang}
\affiliation{ {Department of Physics, Missouri University of Science and Technology, Rolla, MO, US}}
\email{xiaojiewang@mst.edu}

\author{Z.~Wang}
\affiliation{ {Department of Physics, Missouri University of Science and Technology, Rolla, MO, US}}
\email{zhen@umd.edu}

\author[0000-0003-2141-3413]{I.J.~Watson}
\affiliation{{University of Seoul, Seoul, Rep. of Korea}}
\email{ian.james.watson@cern.ch}

\author{S.~Yu}
\affiliation{{Department of Physics, Pennsylvania State University, University Park, PA, USA }}
\email{sjy5345@psu.edu}

\author{X.~Zhang}
\affiliation{{Institute of Nuclear Physics Polish Academy of Sciences, PL-31342 IFJ-PAN, Krakow, Poland }}
\email{xiyingzhangxyz@gmail.com}

\author[0000-0003-0513-3841]{H.~Zhou}
\affiliation{{Tsung-Dao Lee Institute \& School of Physics and Astronomy, Shanghai Jiao Tong University, 800 Dongchuan Rd, Shanghai, SH 200240, China}}
\email{hao_zhou@sjtu.edu.cn}

\collaboration{all}{HAWC Collaboration}

\begin{abstract}

Multi-TeV gamma-ray emission around eHWC J1850+001 (a source from the first HAWC catalog of gamma-ray sources emitting above 56 TeV) is spatially coincident with the pulsar wind nebula (PWN) G32.64+0.53, powered by PSR J1849-0001. The absence of counterparts in radio, optical, and GeV energy ranges, contrasted with clear detections in X-rays and very-high-energy (VHE) gamma-rays, is indicative of a non-thermal leptonic origin for the nebula. We apply a systematic analysis pipeline, including a sophisticated model for the Galactic diffuse emission, to 2860 days of data from the HAWC Observatory. Our detailed analysis confirms that the ultra-high-energy (UHE) emission originates from G32.64+0.53, and we measure its spectrum up to 270 TeV with significant emission well beyond 100 TeV. We fit the multi-wavelength observations with a time-dependent leptonic model powered by the pulsar's rotational energy, and the results establish the nebula as a leptonic PeV accelerator, capable of accelerating electrons to a maximum energy of $E_{\mathrm{cut}}\geq2.9~\mathrm{PeV}$ (95\% one-sided lower bound). The model also constrains the nebular magnetic field to $4.28 ~\mathrm{\mu G}$, supporting a leptonic PWN origin for the observed UHE emission.

\end{abstract}



\section{Introduction}

Ultra-high-energy (UHE) gamma-ray astronomy is providing an unprecedented view into the most powerful particle accelerators within our Galaxy. The recent discovery of numerous Galactic sources emitting photons with energies beyond hundreds of tera-electronvolt (TeV) has confirmed that these objects can accelerate particles to peta-electronvolt (PeV) energies, ushering in the era of UHE astronomy \citep{eHWC, UHELHAASO, 1LHAASO, 2eHWC}. 

Pulsar wind nebulae (PWNe) are nebulae inflated by magnetized, relativistic electrons and positrons (collectively electrons hereafter) winds, powered by the immense rotational energy of rapidly spinning neutron stars. 
These relativistic electrons generate synchrotron radiation in the nebula's magnetic field, seen from radio to X-rays. These electrons also upscatter ambient photons (e.g., cosmic microwave background (CMB) photons) to the very-high-energy (VHE) regime via inverse Compton scattering (ICS).
According to the first catalog of the Large High Altitude Air Shower Observatory (LHAASO), among the 43 UHE sources discovered by LHAASO ($4\sigma$ detection at $E>100~\mathrm{TeV}$), 22 sources are found to be pulsar-associated, including PWNe and pulsar halos \citep{1LHAASO}. This suggests that PWNe are a major class of Galactic accelerators emitting UHE gamma rays through ICS. 

A prime example of such an extreme accelerator is the PWN powered by PSR~J1849-0001, G32.64+0.53, which might accelerate particles to the PeV regime. 
The pulsar itself is highly energetic, with a spin-down luminosity of $\dot{E} = 9.8 \times 10^{36}$~erg~s$^{-1}$ \citep{ATNF}. The pulsar has a characteristic age of $\tau_c=43~\mathrm{kyr}$, placing it in the class of middle-aged pulsars \citep{ATNF, XrayPWN}. Based on the large hydrogen column density derived from X-ray observations, \citet{RXTE} obtained a distance of $7~\mathrm{kpc}$ to the pulsar. 
This distance is also adopted for the modeling in this work.

Previous studies using the High Energy Stereoscopic System (H.E.S.S.) have firmly identified the associated nebula, HESS~J1849-000, as a PWN \citep{RXTE, HGPS, XrayPWN}. Observations in X-rays and VHE gamma-rays can be explained by a leptonic scenario, where a population of accelerated electrons produces the emission \citep{XrayPWN}. The nebula has a compact X-ray core surrounded by a more extended VHE gamma-ray emission \citep{XrayPWN}. This is understood as a consequence of synchrotron cooling: the highest-energy electrons responsible for X-rays cool rapidly and remain close to the central pulsar, while the lower-energy, longer-lived electrons that produce VHE gamma-rays have time to diffuse outwards.

Gamma-ray emission from xHWC J1848+000 \citep{2eHWC}, the HAWC counterpart to HESS J1849-000, was detected at energies above 177 TeV by the High Altitude Water Cherenkov (HAWC) Observatory. This source is the updated counterpart of the previously reported eHWC J1850+001 \citep{eHWC}, based on the extended dataset and improved event reconstruction \citep{HAWCPass5}.
Using the up-to-date HAWC data, we perform a detailed study of this source, hereafter HAWC~J1849-0000. By combining these new HAWC measurements with existing multi-wavelength data, we perform time-dependent modeling of the leptonic emission powered by the pulsar's rotational energy to constrain the physical properties of the system, including its true age, magnetic field, and the spectrum of the accelerated electrons. 

This paper is structured as follows. In Section~\ref{sec:hawc}, we describe the HAWC observations and our data analysis pipeline. In Section~\ref{sec:results}, we present the results of our analysis, focusing on the properties of HAWC~J1849-0000 and the multi-wavelength modeling. We discuss the implications of our findings in Section~\ref{sec:discussions} and provide our conclusions in Section~\ref{sec:conclusions}.

\section{HAWC Observations and Data Analysis} \label{sec:hawc}

\subsection{The HAWC Observatory}

The HAWC observatory is a wide field-of-view detector sensitive to gamma rays from approximately 300~GeV to hundreds of TeV. It is a particle sampling array that detects secondary particles from extensive air showers (EAS) located at an altitude of 4,100~m near Pico de Orizaba, Mexico. The main array consists of 300 water Cherenkov detectors (WCDs) designed to detect extensive air showers produced by incident gamma rays. Each WCD is equipped with four photomultiplier tubes (PMTs) at the bottom, which collect the Cherenkov light produced as shower particles traverse the purified water. The timing and charge information from the PMTs is used to reconstruct the core location, incoming direction, and energy of the primary gamma ray. Detailed descriptions of the detector can be found in \citet{HAWCNIMA}.

\subsection{Analysis Setup} \label{sec:setup}

\begin{figure}[t]
    \centering
    \includegraphics[width=\columnwidth]{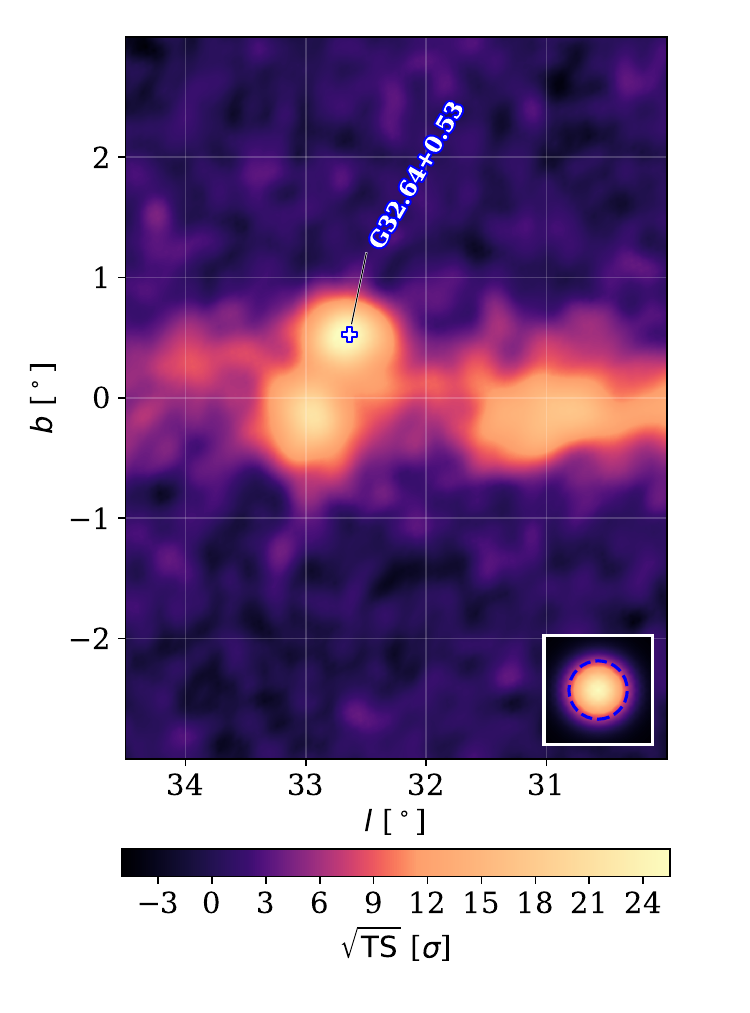}
    \caption{HAWC significance map of the ROI. The map is generated by fitting a test point source with a fixed power-law index ($\alpha=2.5$) at each pixel, optimizing only the flux normalization. 
    The per-pixel test-source fit weights events within a 2$^\circ$ disc by the HAWC PSF (see Section~\ref{sec:setup}).
    The blue label indicates the position of G32.64+0.53 \citep{XrayPWN}.
    The inset in the lower right corner shows a simulated point source injected at the position of HAWC J1849-0000 with the same detector response, assuming a power-law spectrum with a spectral index of $\alpha=2.5$, illustrating the HAWC point spread function (PSF) at this declination band. The blue dashed circle indicates the 68\% containment radius of the PSF.
    }
    \label{fig:hawcmap}
\end{figure}

The analysis presented in this work utilizes 2860 days of HAWC ``Pass~5'' data \citep{HAWCPass5}. We select events with reconstructed energies above 1~TeV, using energies estimated from a neural network \citep{HAWCEE}. Our region of interest (ROI), shown in Figure~\ref{fig:hawcmap}, is defined as a rectangular region spanning between the Galactic longitudes $l \in [30.0^{\circ}, 34.5^{\circ}]$ and latitudes $b \in [-3.0^{\circ}, 3.0^{\circ}]$, which fully contains the source G32.64+0.53.

We perform a binned maximum likelihood analysis using the Python package \texttt{threeML} \citep{threeML} with the \texttt{HAWC Accelerated Likelihood (HAL)} plugin \citep{HAL}. To quantify the preference for a given model (alternative hypothesis, $L_1$) over a baseline model (null hypothesis, $L_0$), we use the test statistic (TS), defined as:
\begin{equation}\label{eq:ts}
    \mathrm{TS} = 2 \times (\ln L_{1} - \ln L_{0}).
\end{equation}

For spectral modeling, we use a power-law (PL) function, defined as:
\begin{equation}\label{eq:pl}
    \Phi(E) = \Phi_{0} \left(\frac{E}{E_{0}}\right)^{-\alpha},
\end{equation}
where $\Phi_{0}$ is the flux normalization at a pivot energy $E_{0}$, and $\alpha$ is the spectral index. 

The significance map shown in Figure~\ref{fig:hawcmap} is produced by fitting the flux normalization $\Phi_0$ of a test point source at each pixel, assuming a PL spectrum with a fixed index of $\alpha=2.5$ and a pivot energy of $E_0=10$~TeV. At every test position, the expected signal in each data pixel within a $2^{\circ}$ disc is weighted by the HAWC point spread function evaluated at its offset from the test position, and the per-pixel significance is reported as $\mathrm{sign(TS)} \times \sqrt{\mathrm{\left|TS\right|}}$, which approximates the pre-trial significance via Wilks' theorem \citep{Wilks}.  Negative values arise when the best-fit normalization is negative. Because the discs of adjacent test positions overlap, the resulting map is spatially correlated on the scale of the HAWC PSF. The map therefore serves as a visualization aid, while quantitative source detection and characterization are performed within the joint multi-source likelihood framework described in Section~\ref{sec:alps}.

To accurately model the complex emission in the ROI, we first establish a physics-based model of the Galactic diffuse emission (GDE; Section~\ref{sec:gde}) and then apply an iterative analysis pipeline to search for individual sources (Section~\ref{sec:alps}).

\subsection{Galactic Diffuse Emission Modeling} \label{sec:gde}

The Galactic plane is a bright, complex source of diffuse gamma rays originating from the interaction of cosmic rays with interstellar gas and radiation fields. To properly model the contribution of individual sources in the ROI, this GDE must be accurately accounted for. We have adopted a sophisticated, energy-dependent GDE template generated with the HERMES software \citep{HERMES}.

This GDE template incorporates emission from ICS, bremsstrahlung, and neutral pion ($\pi^0$) decay from cosmic-ray interactions with atomic (HI) and molecular (H$_2$) hydrogen. The HI and H$_2$ gas maps used as spatial templates are from \citet{HI4PI} and \citet{CfA}, respectively. For the $\pi^{0}$ decay channel, which dominates in the HAWC energy range, we have adopted the cross-section calculations from \citet{Kelner}, as they are valid for proton energies well into the PeV scale. A free scaling parameter was applied to the GDE template within the ROI, to allow its overall contribution to be adjusted in the likelihood fit, accounting for potential local variations in the cosmic-ray sea density. 

\subsection{Source Finding Algorithm} \label{sec:alps}

With the GDE template established as a baseline, we have applied an iterative pipeline to search for and characterize individual gamma-ray sources. This process is adapted from the procedure used for the Boomerang region analysis by the HAWC collaboration \citep{Boomerang}. There is an additional step to test the elliptical morphology of the found extended sources compared to \citet{Boomerang}. The pipeline systematically executes the following steps:

\paragraph{1. Point Source Search}
The pipeline begins by identifying the location of the most significant hotspot in the current residual map. A new test point source is placed at this location and modeled with a PL spectrum (Equation~\ref{eq:pl}), where the positional parameters RA, Dec, the flux normalization ($\Phi_0$) and the spectral index ($\alpha$) are treated as free parameters. The spectral parameters of all sources in the current model are also free parameters. If the inclusion of this source results in a TS value greater than 25, it is added to the source model. This iterative process of identifying hotspots and adding sources is repeated until no new location in the ROI meets the detection threshold.

\paragraph{2. Extension Search}
After the point source search is complete, each source in the model is tested for spatial extension, proceeding iteratively in order of decreasing TS. For each source, its point-like morphology is temporarily replaced with a spatially extended template, typically a radially symmetric Gaussian distribution, with its spectrum, position, and extension size as free parameters. For the remaining sources in the model, their positional parameters (RA, Dec) are fixed while their angular sizes and spectral parameters are left free. We remark that when a morphological model is fitted, the point spread function is convolved with the morphological model. If the extended model is favored over the point-source model with a TS value greater than 16, the source's morphology is updated to the extended model. Modeling a bright source as extended can significantly alter the flux distribution in the region. Therefore, after a source's morphology is updated, all other sources in the model are re-evaluated. Any source whose significance drops below the detection threshold ($\mathrm{TS} < 25$) is removed from the list. After this pruning, all parameters of the remaining sources in the model are refit.

\paragraph{3. Elliptical Morphology Search}
For sources that are found to be extended in the previous step, we perform an additional test for elliptical morphology. This step was not included in the original pipeline presented in \citet{Boomerang}. The radially symmetric Gaussian morphology is replaced with an elliptical 2D Gaussian model, which includes two additional free parameters: the eccentricity $e$ and the position angle $\theta$ (see Appendix~\ref{sec:specmorph}). The elliptical model is adopted if it is statistically preferred over the symmetric one with a $\mathrm{TS} > 16$. 
During the elliptical fitting step, the source under test is described by an elliptical 2D Gaussian with five free morphological parameters (RA, Dec, $\sigma$, $e$, $\theta$) together with its spectral model. For all other sources in the model, the central coordinates (RA, Dec) are held fixed at their best-fit values from the previous iteration, while their angular sizes and spectral parameters are left free to readjust.

Detailed descriptions of the spectral and morphological models used in this analysis are presented in Appendix~\ref{sec:specmorph}. The results of this source-finding procedure, including the composition of the final best-fit model for the ROI, are presented in Section~\ref{sec:results}.

\section{Results} \label{sec:results}

\subsection{Source Search Results}

Applying the source-finding algorithm described in Section~\ref{sec:alps}, our analysis results in a final best-fit model for the ROI that consists of the GDE component plus four individual sources: three bright, extended sources and one additional point-like source. One of the extended sources is found to prefer the elliptic morphology.

The initial application of the pipeline resolved the emission in the vicinity of PSR~J1849-0001 into two nearby point sources with an angular separation of only $\sim0.2^{\circ}$. We evaluate this preference using the Bayesian Information Criterion (BIC) \citep{BIC}. We adopt BIC over the Akaike Information Criterion (AIC) because it imposes a stricter penalty on additional parameters, helping prevent overfitting. The single extended source model is favored by a $\Delta\text{BIC} = -49$ \citep{BIC}. This result is consistent with previous findings from H.E.S.S. \citep{HGPS}.

{
\begin{deluxetable}{l|ll|CC}[ht!]
\tablewidth{0pt}
\tabletypesize{\footnotesize}
\tablecaption{Summary for the spectral and morphological model tests \label{tab:source_model}}
\tablehead{
\colhead{Source} & \multicolumn{1}{|c}{Spectrum} & \colhead{Morphology} & \multicolumn{1}{|c}{$N_ \mathrm{par}$} & \colhead{$\Delta \mathrm{BIC}$}}
\renewcommand{\arraystretch}{1.4}
\startdata
J1848-0146 & LogP & Elliptical & 8 & $-18$ \\
\textbf{J1848-0146} & \textbf{COPL} & \textbf{Elliptical} & 8 & $-30$ \\
\hline
J1849-0000 & LogP & Gaussian & 6 & $-15$ \\
J1849-0000 & COPL & Gaussian & 6 & $-7$ \\
J1849-0000 & PL & Laplace & 5 & $-4$ \\
\textbf{J1849-0000} & \textbf{LogP} & \textbf{Laplace} & 6 & $-23$ \\
J1849-0000 & COPL & Laplace & 6 & $-14$ \\
\hline
J1852-0002 & LogP & Gaussian & 6 & $-54$ \\
J1852-0002 & COPL & Gaussian & 6 & $-53$ \\
J1852-0002 & PL & Laplace & 5 & $-9$ \\
\textbf{J1852-0002} & \textbf{LogP} & \textbf{Laplace} & 6 & $-72$ \\
J1852-0002 & COPL & Laplace & 6 & $-70$ \\
\hline
\enddata
\end{deluxetable}}

For the three extended sources, we have tested alternative spectral and morphological models to find the best description for each. We have tested two additional spectral models, a power-law with an exponential cutoff (COPL) and a log-parabola (LogP), against the baseline PL model. The COPL model extends the simple power-law with a high-energy exponential cutoff term. It is defined as:
\begin{equation}\label{eq:copl}
    \Phi(E) = \Phi_{0} \left(\frac{E}{E_{0}}\right)^{-\alpha} \exp{\left(-\frac{E}{E_{\mathrm{cut}}}\right)},
\end{equation}
where $E_{\mathrm{cut}}$ is the cutoff energy. The LogP model introduces a curvature in the spectrum, which is parabolic in a log-log representation. It is defined as:
\begin{equation}\label{eq:logp}
    \Phi(E) = \Phi_{0} \left(\frac{E}{E_{0}}\right)^{-\alpha - \beta \log_{10}(E/E_{0})},
\end{equation}
where $\beta$ is the spectral curvature parameter. 

For the two sources with initially symmetric morphologies, we have also tested a 2D Laplace distribution against the baseline Gaussian model. The definitions of these models are presented in Appendix~\ref{sec:specmorph}.

The results of these model comparison tests are summarized in Table~\ref{tab:source_model}. For each test, we report $\Delta\text{BIC}$ relative to the baseline model where all sources have a PL spectrum and a Gaussian morphology. By selecting the model with the lowest $\Delta\text{BIC}$ for each source, we have determined the optimal model configuration based on the criterion of minimizing the BIC. $N_{\mathrm{par}}$ indicates the number of free parameter for each source's model. As shown in the table, the source with an intrinsically elliptical morphology (J1848-0146) is best described by a COPL spectrum. The other two extended sources (J1849-0000 and J1852-0002) are both best fitted by a LogP spectrum combined with a Laplace spatial distribution.

\begin{figure*}[t]
    \centering
    \gridline{\fig{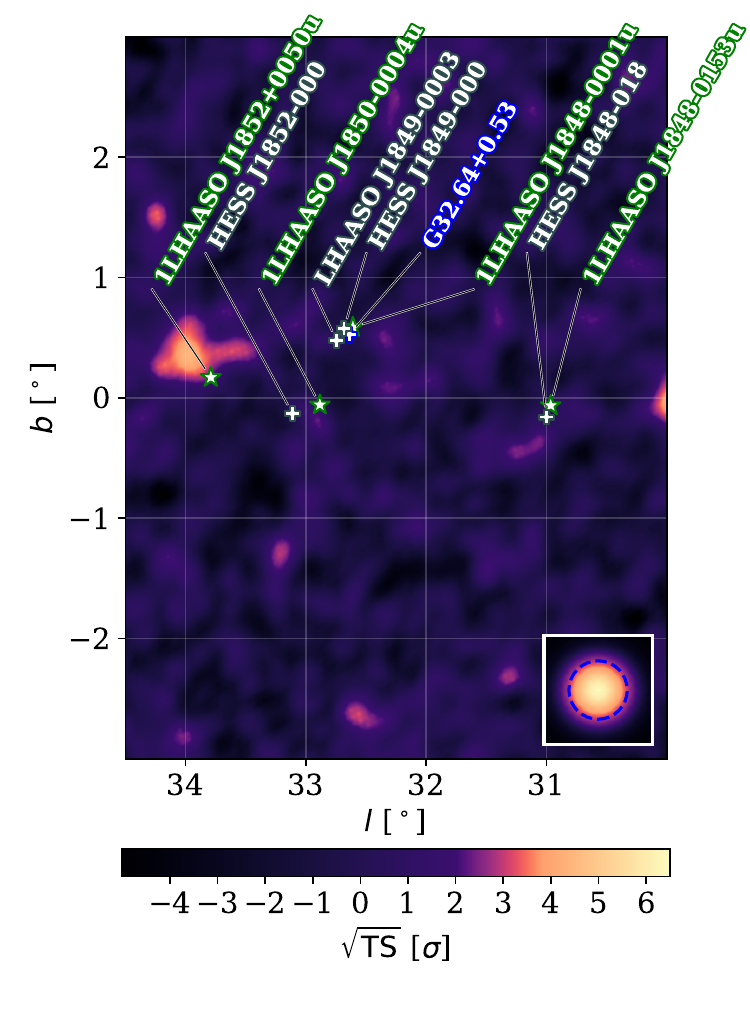}{0.4\textwidth}{(a) Residual map}
          \fig{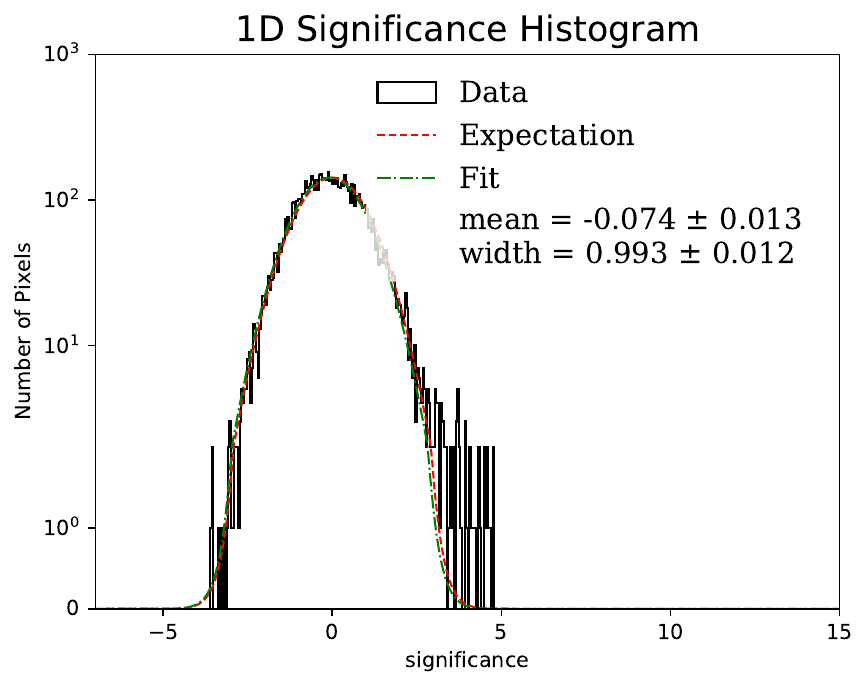}{0.58\textwidth}{(b) 1D histogram of the residual map}}
    \caption{(a) Residual significance map after modeling the three extended sources, but before adding the final point source. The upper green labels indicate the positions of the sources measured by LHAASO KM2A, published at the first LHAASO catalog \citep{1LHAASO}. The other labels are taken from TeVCat \citep{TeVCat}. A hotspot with a significance of $\sim 4.5\sigma$ is visible at $(l,b) \approx (33.96^{\circ}, 0.36^{\circ})$, which is close to 1LHAASO~J1852+0050u. (b) The one-dimensional version of (a), with a positive tail extending to around $4\sigma$. Because the underlying map is spatially correlated on the scale of the HAWC PSF (Section~\ref{sec:setup}), this histogram is presented only as a qualitative visualization of the per-pixel significance distribution and its high-significance tails.}
    \label{fig:residual_beforepsadd}
\end{figure*}

After optimizing the models for these three extended sources, the GDE contribution in the region around $(l,b)=(33.96^{\circ}, 0.36^{\circ})$ were reduced compared to the initial pipeline result. We identified a region of residual emission with a significance of $\sim 4.5\sigma$, as shown in Figure~\ref{fig:residual_beforepsadd}. This emission is spatially coincident with 3HWC J1852+013 and 1LHAASO J1852+0050u, as shown in Table~\ref{tab:correspondings}. Given these counterparts and the significant residual, we tested the hypothesis of an additional point source at this hotspot, and detected a new source with a $\mathrm{TS} = 25.2$, exceeding the threshold value of 25. Thus, we included this fourth source in our model. The final, complete model for the ROI is obtained by performing a final fit of all the parameters for all four sources. We verified that the inclusion of this additional source has a negligible impact on the best-fit parameters of the main source, HAWC J1849-0000, with variations remaining well within statistical uncertainties.

\begin{deluxetable}{c|c}[ht]
    \tabletypesize{\small}
    \tablecaption{Best-fit values of HAWC~J1849-0000 \label{tab:j1849bestfit}}
    \tablehead{ \colhead{Parameter} & \colhead{Best-fit value} }
    \startdata
    RA [deg] & $282.26^{+0.01}_{-0.01}$ \\ 
    Dec [deg] & $0.00^{+0.01}_{-0.01}$ \\ 
    Size [deg] & $0.09^{+0.01}_{-0.01}$ \\ 
    $\Phi_{0}$ [$\mathrm{(TeV~cm^{2}~s})^{-1}$] & $9.37^{+1.11+1.19}_{-0.99-1.19}\times10^{-15}$ \\
    $\alpha$ & $2.07^{+0.11+0.07}_{-0.11-0.07}$ \\
    $\beta$ & $0.19^{+0.06+0.03}_{-0.06-0.03}$ \\
    TS & 347 \\
    \enddata
\tablecomments{The first set of errors is statistical, while the second is systematic. $N_{\mathrm{par}}$ is 6.}
\tablecomments{RA is right ascension and Dec is declination. }
\end{deluxetable}

\begin{figure*}[ht!]
\centering
\gridline{\fig{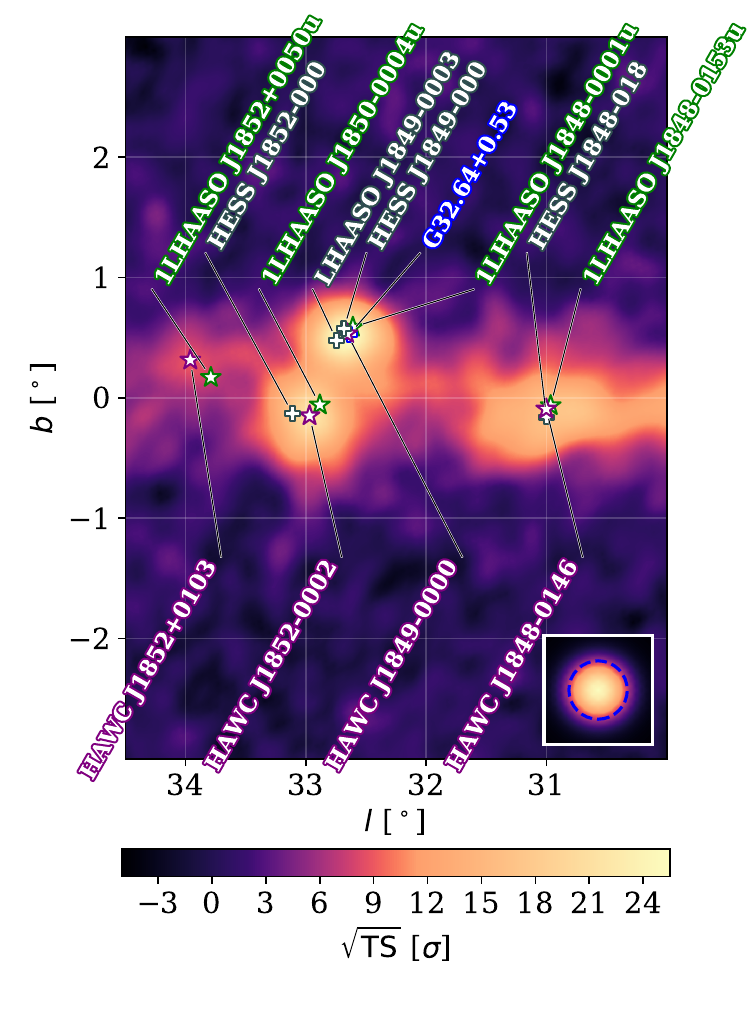}{0.3\textwidth}{(a) HAWC significance map}
          \fig{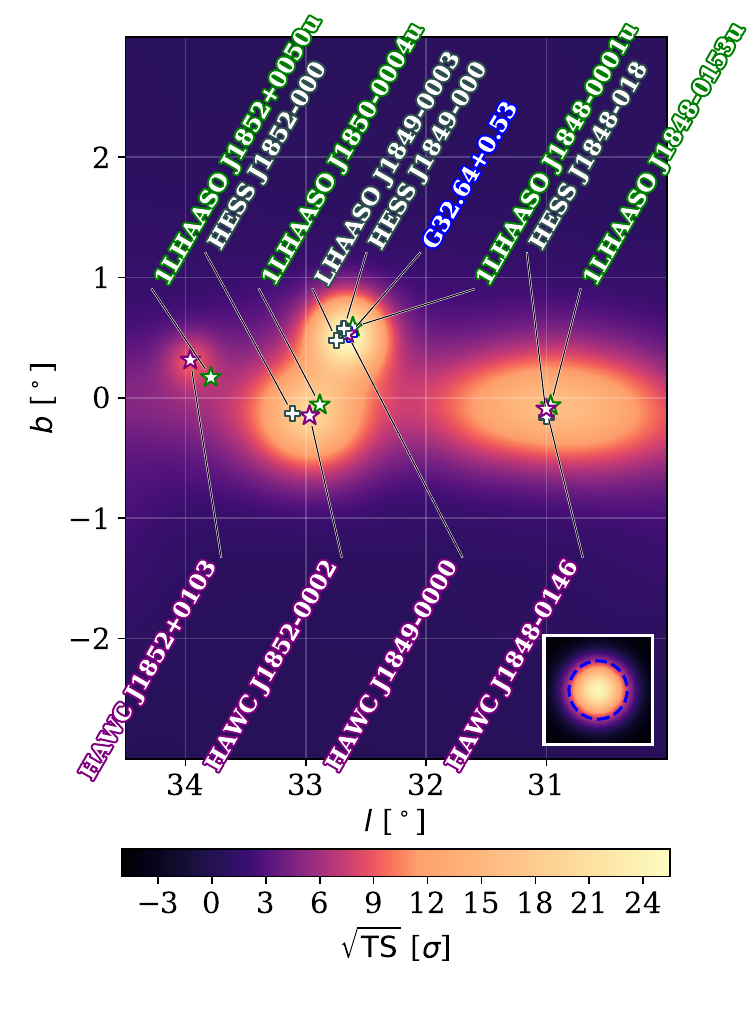}{0.3\textwidth}{(b) Model map}
          \fig{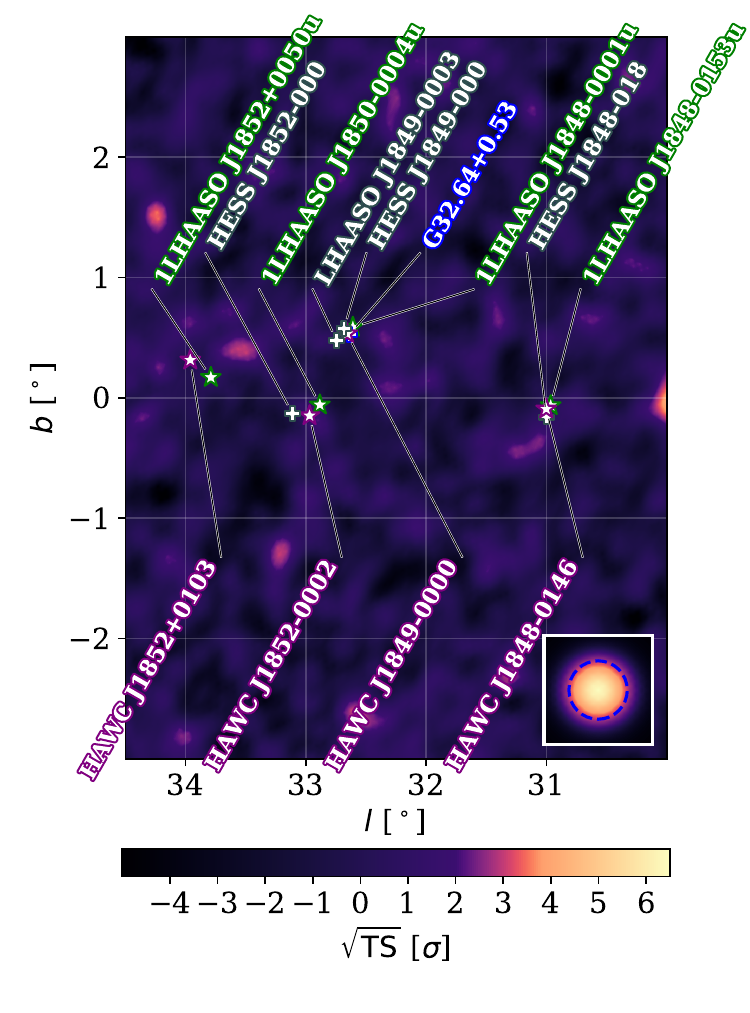}{0.3\textwidth}{(c) Residual map}}
\gridline{\fig{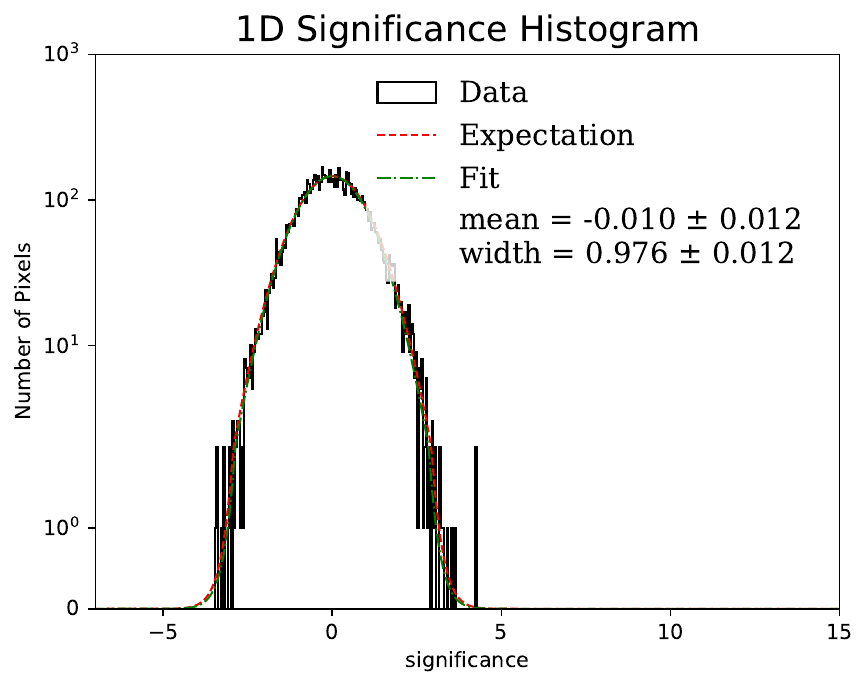}{0.4\textwidth}{(d) 1D histogram of the residual map}}
\caption{Sky maps of the ROI in Galactic coordinates. (a) The initial HAWC significance map. (b) The significance map produced from our final, four-source model. The purple labels indicate the sources found and modeled in this work. (c) The final residual significance map. (d) The one-dimensional version of (c), showing no significant emission remains after subtracting our model from the data. The two pixels around $4\sigma$ are from the edge of the ROI at $l\sim30\degr$.}\label{fig:sigmaps}
\end{figure*}

Figure~\ref{fig:sigmaps} shows the initial HAWC significance map (a), the map of our final best-fit model (b), and the final residual map (c). Panel (d) of Figure~\ref{fig:sigmaps} is the one-dimensionally binned version of the residual map (c), which demonstrates that the residual map is devoid of any significant excess, confirming the goodness-of-fit of our model.  
The best-fit parameters and TS of HAWC~J1849-0000, the source of interest, are presented in Table~\ref{tab:j1849bestfit}, while the best-fit parameters of the other three sources are summarized in Table~\ref{tab:fitresult}. 
All sources are detected with a significance well above our detection threshold of $\sqrt{\mathrm{TS}} > 5$.

The four sources identified in this work have positional counterparts in existing VHE catalogs. These counterparts are listed in Table~\ref{tab:correspondings}.

\begin{deluxetable*}{CCCC}[ht]
    \tablecaption{VHE counterparts for the sources found in this work. LHAASO KM2A sources are from the first LHAASO catalog \citep{1LHAASO}; H.E.S.S. sources are from the H.E.S.S. Galactic Plane Survey \citep{HGPS}. The third HAWC catalog found three sources in the region \citep{3HWC}. \label{tab:correspondings}}
    \tablehead{ \colhead{HAWC (This work)} & \colhead{1LHAASO (KM2A)} & \colhead{H.E.S.S.} & \colhead{HAWC (3HWC)} }
    \startdata
    \text{J1848-0146} & \text{J1848-0153u} & \text{J1848-018} & \text{J1847-017} \\
    \text{J1849-0000} & \text{J1848-0001u} & \text{J1849-000} & \text{J1849+001} \\
    \text{J1852-0002} & \text{J1850-0001u} & \text{J1852-000} & \nodata \\
    \text{J1852+0103} & \text{J1852+0050u} & \nodata & \text{J1852+013} \\
    \enddata
\end{deluxetable*}

The source HAWC~J1848-0146 is found to have an elliptical morphology with an eccentricity of $e = 0.93\pm0.02$. This value is consistent with the reported eccentricity of its LHAASO counterpart, 1LHAASO~J1848-0153u, which was measured to be $0.92$ (KM2A) and $0.95$ (WCDA) \citep{LHAASOW43}. The point source HAWC~J1852+0103 was detected in the third HAWC catalog \citep{3HWC}. The source was not detected by H.E.S.S. but is positionally coincident with the 1LHAASO~J1852+0050u.

\begin{figure}[ht!]
    \centering
    \includegraphics[width=\columnwidth]{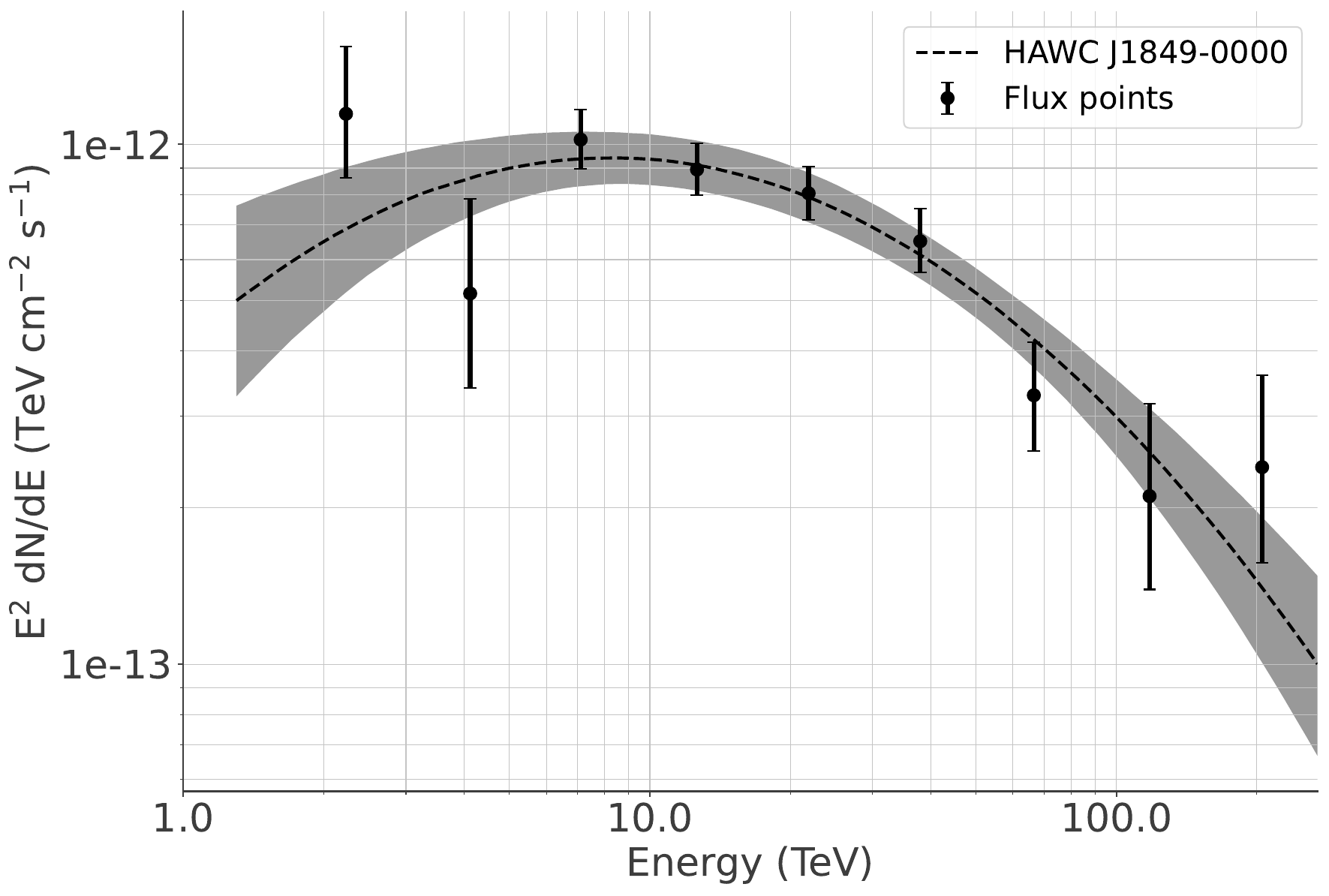}
    \caption{The spectral energy distribution (SED) of HAWC~J1849-0000. 
    The data points represent the flux in each energy bin, with error bars showing $1\sigma$ statistical uncertainties. 
    The dashed line is the best-fit spectral model (Log-Parabola), and the shaded region is the $1\sigma$ confidence interval. 
}
    \label{fig:hawcsed}
\end{figure}

This work focuses on the most statistically significant source in the ROI, HAWC~J1849-0000, which is identified as PWN G32.64+0.53 based on its spatial coincidence with the X-ray and VHE gamma-ray counterparts and its association with PSR~J1849-0001 \citep{XrayPWN, HGPS}. The spectral energy distribution (SED) for this source is presented in Figure~\ref{fig:hawcsed}, and the corresponding flux points are listed in Table~\ref{tab:fluxpoints}. The flux points for the SED are calculated by fitting the flux normalization $\Phi_{0}$ in each energy bin while keeping the other spectral parameters of the global best-fit model fixed, as described in \citet{HAWCEE}. Note that the horizontal position of each flux point represents the logarithmic center of the reconstructed energy bin. Due to energy dispersion, the true energy distributions of events in adjacent bins overlap. While this introduces correlations between the flux points, the spectral parameters reported in this work are obtained through a forward-folding likelihood analysis, which accounts for energy dispersion. The source is detected across nine energy bins from 1.3~TeV to 270~TeV, extending well beyond 100~TeV with a significance of $\sqrt{\mathrm{TS}} \approx 4.7$ in the highest energy bin. This confirms the source's status as a powerful UHE accelerator. 

\begin{deluxetable}{CCC}[ht]
    \tablecaption{Flux Points of HAWC~J1849-0000 \label{tab:fluxpoints}}
    \tablehead{ \colhead{Energy [TeV]} & \colhead{Flux [$10^{-13}~\mathrm{TeV\,cm^{-2}\,s^{-1}}$]} & \colhead{$\mathrm{\sqrt{TS}}$}}
    \startdata
    $2.2$ &   $11.46_{-2.80}^{+3.94}$ & 3.4 \\
    $4.1$ &   $5.11_{-1.79}^{+2.90}$ & 2.4 \\
    $7.1$ &   $10.22_{-1.23}^{+1.45}$ & 7.2 \\
    $12.6$ &  $8.96_{-0.98}^{+1.08}$ & 10.3 \\
    $21.9$ &  $8.08_{-0.90}^{+0.99}$ & 11.8 \\
    $38.0$ &  $6.55_{-0.86}^{+0.96}$ & 10.8 \\
    $66.5$ &  $3.29_{-0.71}^{+0.91}$ & 7.0 \\
    $117.8$ & $2.13_{-0.71}^{+1.00}$ & 4.3 \\
    $205.3$ & $2.38_{-0.79}^{+1.22}$ & 4.7 \\
    \enddata
    \tablecomments{The errors are only statistical.}
\end{deluxetable}

The positional and morphological properties of HAWC~J1849-0000 show agreement with its counterparts at other VHE observatories. The angular separation between the best-fit position from our analysis and the cataloged position of HESS~J1849-000 is only $0.003^\circ$, and the separation from 1LHAASO~J1848-0001u is $0.05^\circ$ \citep{HGPS, 1LHAASO}. Although the Tibet AS$\gamma$ study in \citet{TibetPWN} does not explicitly provide a best-fit position derived from a likelihood fit for this source, it identifies the brightest pixel above 100 TeV at (RA, Dec)=(282.33, 0.08). This location remains consistent with the HESS J1849-000 position within the uncertainty and is also in agreement with the best-fit position of HAWC J1849-0000 determined in this work \citep{TibetPWN}.

\tabletypesize{\scriptsize}
\begin{deluxetable}{cccccccccccc}
    \centerwidetable
    \tablecaption{Best-fit values for the found sources. \label{tab:fitresult}}
    \tablehead{
    \colhead{Source} & \colhead{RA [$\degr$]} & \colhead{Dec [$\degr$]} & \colhead{Size [$\degr$]} & \colhead{$e$} & \colhead{$\theta$ [$\degr$]} & \colhead{$\Phi_{0}$ [$\left(\mathrm{TeV~cm^{2}~s}\right)^{-1}$]} & \colhead{$\alpha$} & \colhead{$\beta$} & \colhead{$E_{\mathrm{cut}}$ [TeV]} & \colhead{$N_{\mathrm{par}}$} & \colhead{TS}
    }
    \startdata
    J1848-0146 & $282.06^{+0.03}_{-0.03}$ & $-1.76^{+0.05}_{-0.05}$ & $0.67^{+0.07}_{-0.07}$ & $0.93^{+0.02}_{-0.02}$ & $65^{+3}_{-3}$ & $3.45^{+0.64+0.79}_{-0.53-0.79}\times10^{-14}$ & $2.16^{+0.13+0.18}_{-0.14-0.18}$ & \nodata & $30^{+9+7}_{-7-7}$ & $8$ & $230$ \\
    J1852-0002 & $283.01^{+0.03}_{-0.03}$ & $-0.04^{+0.02}_{-0.02}$ & $0.23^{+0.02}_{-0.02}$ & \nodata & \nodata & $2.31^{+0.31+0.45}_{-0.27-0.45}\times10^{-14}$ & $2.32^{+0.08+0.09}_{-0.08-0.09}$ & $0.32^{+0.06+0.05}_{-0.06-0.05}$ & \nodata & $6$ & $189$ \\ 
    J1852+0103 & $283.05^{+0.03}_{-0.03}$ & $+1.06^{+0.03}_{-0.03}$ & \nodata & \nodata & \nodata & $1.31^{+0.32+0.15}_{-0.26-0.15}\times10^{-15}$ & $2.73^{+0.17+0.10}_{-0.16-0.10}$ & \nodata & \nodata & $4$ & $25$ \\ 
    GDE & \nodata & \nodata & \nodata & \nodata & \nodata & $5.99^{+0.45+0.79}_{-0.43-0.79}$ & & \nodata \nodata & \nodata & $1$ & $137$ \\ 
    \enddata
\tablecomments{The first set of errors is statistical, while the second is systematic.}
\tablecomments{$\Phi_{0}$ of the GDE is the dimensionless scale factor.}
\end{deluxetable}

Our best-fit source extension of $0.09^\circ$ is consistent with that reported for HESS~J1849-000 \citep{HGPS}. Furthermore, this value is compatible with the constraint from LHAASO that placed an upper limit on the 39\% containment radius ($r_{39}$) of $<0.09^\circ$ \citep{1LHAASO}. This spatial coherence across multiple instruments and energy ranges supports a common physical origin for the observed emission. While the measured size approaches the size of the HAWC point spread function (PSF), the sufficient statistical significance (TS=347) enables the forward-folding likelihood analysis with the PSF-convolved spatial model to statistically distinguish between a pure point source and a morphology with intrinsic extension comparable to the PSF.

\subsection{Systematic Uncertainties}

In addition to the statistical uncertainties, we have estimated the systematic uncertainties for our final model parameters. These uncertainties account for potential discrepancies between simulation and data, as well as known detector biases. The main sources of systematics considered include the late-light effect, the charge uncertainty, the detection threshold of PMT, and variations in PMT efficiencies in time \citep{HAWCEE, HAWCPass5}. 
To ensure a conservative estimate, we selected the maximum absolute deviation observed for each parameter across these systematic sources. These individual maxima were then combined in quadrature to yield the total systematic uncertainty. The resulting values represent a conservative uncertainty and are applied symmetrically to the best-fit parameters, as the second uncertainty term for each parameter in Table~\ref{tab:j1849bestfit},~\ref{tab:fitresult}.

The influence of systematic effects on the morphological parameters (source position and extension) was evaluated and found to be minimal. The estimated systematic variations are smaller than $0.005$, which is at least an order of magnitude smaller than the statistical uncertainties (typically $>0.01$). Therefore, we treat these systematic uncertainties as negligible and do not explicitly list them in the results tables.

\section{Discussion} \label{sec:discussions}

In this section, we interpret the new HAWC measurements within the framework of a multi-wavelength, time-dependent, leptonic model for the pulsar wind nebula powered by the rotation of PSR~J1849-0001. This model assumes electrons are continuously injected from the pulsar and lose energy via synchrotron radiation and ICS \citep{PWNEvolution}. The model framework is adapted from methods previously used to describe other detected PWNe, such as HAWC~J2019+368 and HAWC~J1809-1919 \citep{HAWCJ2019, HAWCJ1809}. 

The energy source for the nebula is the rotational power of the central pulsar, known as the spin-down luminosity: 
\begin{equation}
    \dot{E} = \frac{4\pi^{2}I \dot{P}}{P^{3}}.
\end{equation}
For PSR~J1849-0001, the observed period $P = 38.5~\mathrm{ms}$ and its time derivative $\dot{P} = 1.4156\times10^{-14}~\mathrm{s~s^{-1}}$ yield a present-day spin-down luminosity of $\dot{E} \approx 9.8~\times10^{36}~\mathrm{erg~s^{-1}}$, assuming a canonical neutron star moment of inertia $I = 10^{45}~\mathrm{g~cm^{2}}$ \citep{PWNEvolution}.

Under the assumption that the spin-down luminosity is used for magnetic dipole radiation, $\dot{\Omega}\propto \Omega^{3}$ where $\Omega=2\pi/P$, the spin-down luminosity evolves with time $t$ as:
\begin{equation}
    \dot{E}(t) = \dot{E}_{0}\left(1+\frac{t}{\tau_{0}}\right)^{-2},
\end{equation}
where $\dot{E}_{0}$ is the initial spin-down luminosity and $\tau_{0}=P_{0}^{2}/(2\dot{P}P)$ is the spin-down timescale. This timescale is determined by the pulsar's initial spin period, $P_0$. Since $P$ and $\dot{P}$ are measured, $P_{0}$ determines the true age $\tau$ of the system through Equation~\ref{eq:age} below. 

We assume that a fraction of the instantaneous spin-down power, $\eta$, is converted into a population of relativistic electrons. This conversion efficiency $\eta$ is treated as a free parameter. The injected electron spectrum is assumed to follow COPL in Equation~\ref{eq:copl} with $E_0=1~\mathrm{TeV}$.  
These constitute free parameters of our model.

Under the same assumption, the age of the pulsar can be estimated:
\begin{equation}\label{eq:age}
    \tau = \frac{P}{2\dot{P}}\left[1-\left(\frac{P_{0}}{P}\right)^{2}\right].
\end{equation}
If $P_{0} \ll P$, $\tau$ becomes a characteristic age $\tau_{c}$, which is $43~\mathrm{kyr}$ for PSR~J1849-0001. We adopt the age $\tau=9~\mathrm{kyr}$ estimated for this system from the empirical TeV-to-X-ray flux-ratio relation~\citep{PWNAge, XrayPWN}, which through this relation fixes the initial spin period to $P_{0}=34.2~\mathrm{ms}$.

The magnetic field within the nebula is also assumed to evolve over time. Its evolution can be parameterized as:
\begin{equation}
    B(t) = B_{0}\left(1+\sqrt{\frac{t}{\tau_{0}}}\right)^{-1},
\end{equation}
where $B_0$ is the initial magnetic field strength \citep{HAWCJ2019}. In our model, we directly fit for the magnetic field strength at the pulsar's current age, $\tau$. Thus, the current magnetic field $B(\tau)$ serves as a free parameter.

The two zones are defined as follows. The relic zone consists of electrons injected and evolved over the full true age of the system, $[0,\ \tau]$, and primarily accounts for the gamma-ray emission. The young zone consists of electrons injected and evolved only over the most recent $t_{\mathrm{frac}} \times \tau$, i.e., from $\tau(1 - t_{\mathrm{frac}})$ to $\tau$, and primarily accounts for the X-ray synchrotron emission. The fraction $t_{\mathrm{frac}}$ is treated as a free parameter, with smaller values corresponding to a younger, more compact electron population responsible for the X-ray emission. 

Our time-dependent leptonic model therefore has five free parameters: the conversion efficiency $\eta$, the electron spectral index $\alpha$, the cutoff energy $E_{\mathrm{cut}}$, the current magnetic field strength $B(\tau)$, and the age fraction $t_{\mathrm{frac}}$. We implement the model using the GAMERA software package \citep{GAMERA} and fit the multi-wavelength data using a Bayesian approach with the \texttt{emcee} Markov Chain Monte Carlo (MCMC) framework \citep{emcee}. We assess convergence with the rank-normalized potential scale reduction factor $\hat{R}$ \citep{RubinStatistic}, obtaining $\hat{R}<1.01$ for all parameters with effective sample sizes exceeding $10^{6}$.

\begin{figure*}[ht]
    \centering
    \gridline{\fig{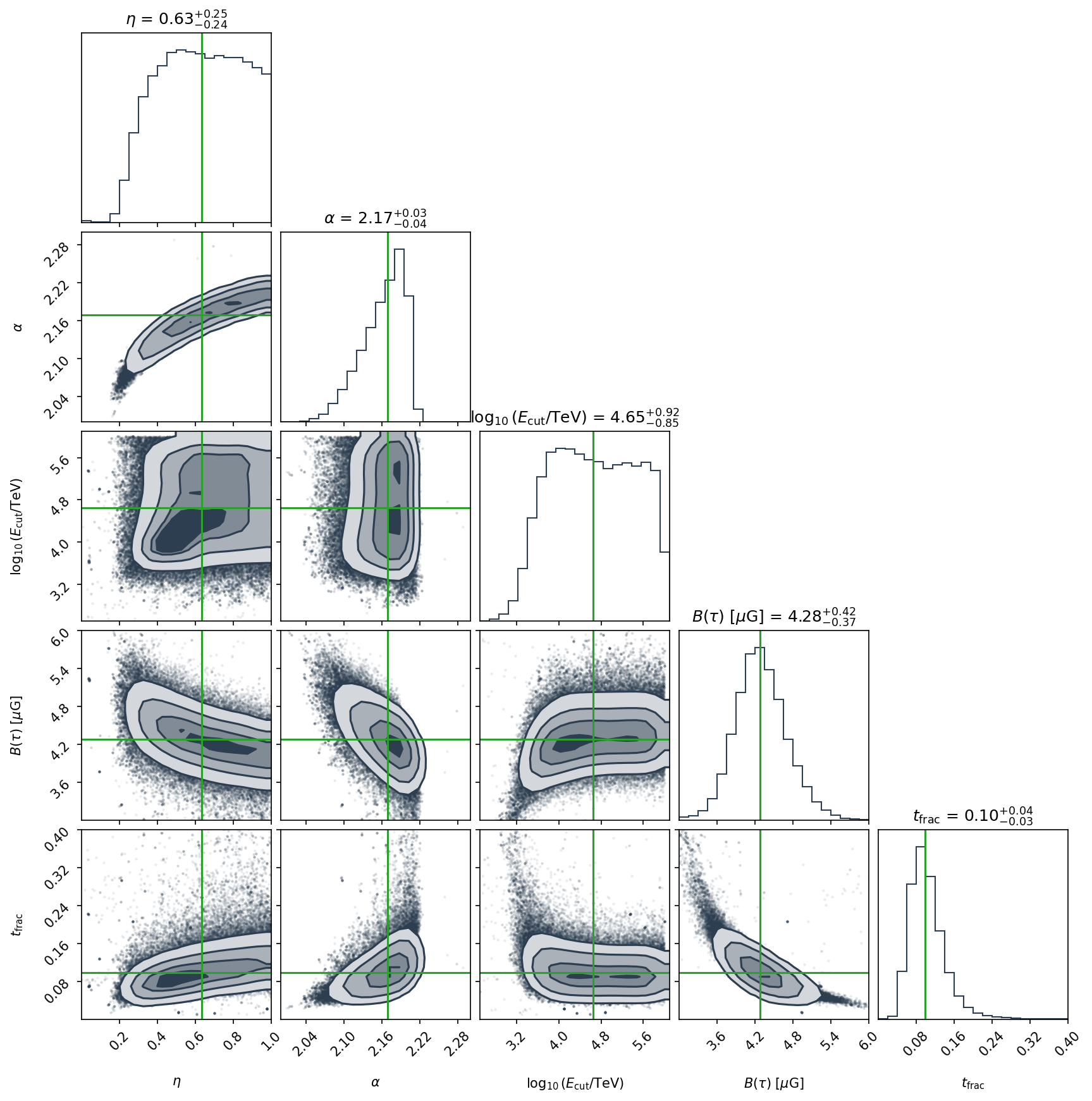}{0.4\textwidth}{(a) Posterior distributions}
          \fig{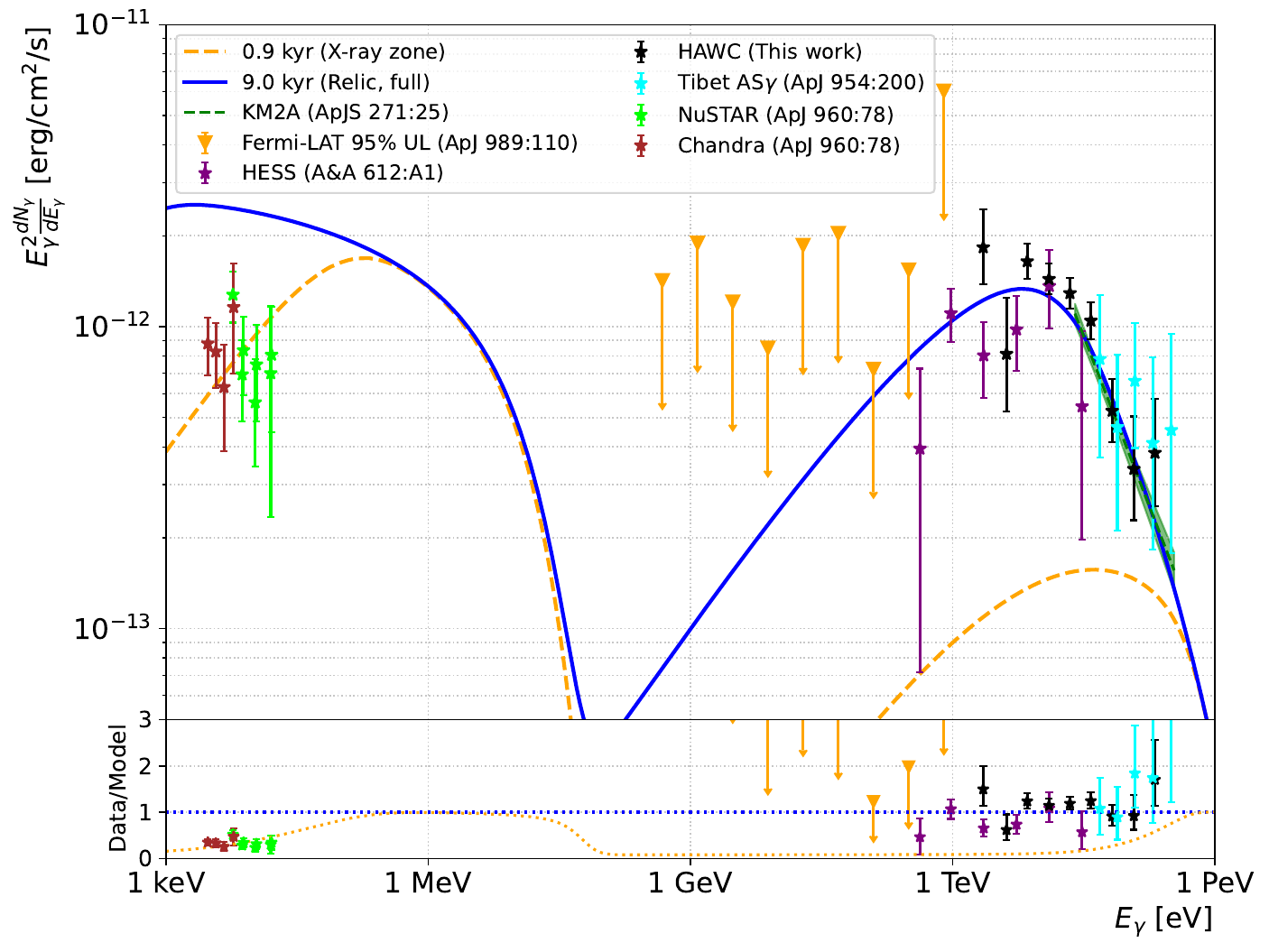}{0.59\textwidth}{(b) The best-fit SED}}
    \caption{(a) The posterior distributions obtained by MCMC for the two-zone, time-dependent leptonic model. The diagonal panels show the marginalized posterior distributions for $\eta$, $\alpha$, $\log_{10}\left(E_{\mathrm{cut}}/\mathrm{TeV}\right)$, $B\left(\tau\right)$, and $t_{\mathrm{frac}}$, where $t_{\mathrm{frac}}$ is the fraction of the true age defining the duration of the young electron population. The green lines indicate the median values of the distributions. (b) The best-fit SED from the two-zone model, overlaid with multi-wavelength data from Chandra, NuSTAR, Fermi-LAT, H.E.S.S., LHAASO, Tibet-AS$\gamma$, and this work (HAWC) \citep{XrayPWN, FermiPWN, HGPS, 1LHAASO, TibetPWN}. The blue solid curve represents the relic zone (electrons evolved over the full age $\tau$), which accounts for the gamma-ray emission, while the orange dashed curve represents the young zone (electrons evolved over $t_{\mathrm{frac}} \times \tau$), which accounts for the X-ray synchrotron emission.}
    \label{fig:leptonicsed2}
\end{figure*}

\begin{deluxetable}{CC}[ht]
    \tablecaption{Best-fit parameters of the two-zone leptonic model. Values are posterior medians with 68\% credible intervals. The age is fixed to $\tau=9~\mathrm{kyr}$ \citep{PWNAge, XrayPWN}. \label{tab:leptonic2}} 
    \tablehead{ \colhead{Parameter} & \colhead{Median (68\% CI)} }
    \startdata
    \text{Conversion Efficiency,}~$\eta$ & $0.63_{-0.24}^{+0.25}$ \\
    \text{Electron Spectral Index,}~$\alpha$ & $2.17_{-0.04}^{+0.03}$ \\
    \text{Electron Cutoff Energy,}~$\log_{10}\left(E_{\mathrm{cut}}/\mathrm{TeV}\right)$ & $4.65_{-0.85}^{+0.92}$ \\
    \text{Current B-Field,}~$B(\tau)~[\mathrm{\mu G}]$ & $4.28_{-0.37}^{+0.42}$ \\
    \text{Age Fraction,}~$t_{\mathrm{frac}}$ & $0.10_{-0.03}^{+0.04}$ 
    \enddata
\end{deluxetable}

The best-fit SED and posterior distributions of the two-zone model are shown in Figure~\ref{fig:leptonicsed2}, and the best-fit parameters are summarized in Table~\ref{tab:leptonic2}. The two-zone model provides a good description of the data across the full energy range ($\chi^{2}/\mathrm{dof}=1.02$). With the age fixed to $\tau=9~\mathrm{kyr}$, the inferred age fraction $t_{\mathrm{frac}}=0.10$ corresponds to a young-electron injection over the most recent $0.9~\mathrm{kyr}$. This is consistent with the compact X-ray core being more recent and spatially smaller than the extended gamma-ray emission.

The conversion efficiency is not tightly constrained by the current data. The marginalized posterior for $\eta$ is broad and asymmetric, peaking near $\eta\approx0.5$ with a median of $0.63$ ($68\%$ credible interval (CI) $0.39–0.88$), and it remains substantial up to the upper prior boundary $\eta=1$ (Figure~\ref{fig:leptonicsed2}a), with a $95\%$ CI of $0.25–0.98$. The data therefore favor an efficient conversion of the spin-down power into relativistic electrons, but do not determine $\eta$ precisely and do not exclude values close to $\eta=1$.

For comparison, \citet{XrayPWN} do not fit the efficiency but adopt a large value, $\eta\approx0.95$, motivated by the non-detection of PSR~J1849-0001 by the Fermi-LAT. They further verified that different values of $\eta$ could be readily accommodated in their model by adjusting other parameters such as the age, magnetic field, seed-photon density, and injection index \citep[][their Section~3.3]{XrayPWN}. This non-uniqueness is expected given the covariance among broadband PWN model parameters \citep{Park2023}, and the breadth of our posterior reflects the same degeneracy: the present multi-wavelength data neither require nor exclude $\eta\approx0.95$. A closely analogous multi-zone analysis of HESS~J1809$-$193, hosting the pulsar PSR J1809-1917 with a similar characteristic age($\sim 51$ kyr) as that of PSR J1849-0001, with the same time-dependent framework infers $\eta\approx0.67$ \citep{HAWCJ1809}, similar to the value found here. The conversion efficiency can vary substantially among PWNe for the chosen theoretical model, for example, a leptonic model of the Crab Nebula ($\tau_{\mathrm{c}} = 1.2\ {\mathrm{kyr}}$) infers $\eta = 0.265 \pm 0.0032$ \citep{Nie_2022} and a leptonic model for 1LHAASO J1945+2424 assuming a pulsar with an age of $\tau_{\mathrm{c}} = 10\ {\mathrm{kyr}}$ infers $\eta = 0.2$ \citep{Araya2024}.

The electron cutoff energy is constrained to $E_{\mathrm{cut}}\geq2.9~\mathrm{PeV}$ (95\% one-sided lower bound), which establishes HAWC~J1849-0000 as a PeV electron accelerator. In the Klein-Nishina regime, these PeV electrons can produce gamma rays reaching approximately $1.5~\mathrm{PeV}$ via ICS on CMB photons, following the relation $E_{e} \simeq 2.15\left(E_{\gamma}/1~\mathrm{PeV}\right)^{0.77}~\mathrm{PeV}$ \citep{UHELHAASOCrab}.

The recent injection of the young electron population is consistent with the picture in which the highest-energy electrons responsible for the X-ray synchrotron emission cool rapidly and remain close to the pulsar, while the longer-lived population producing the inverse-Compton gamma rays traces the full age of the nebula. Future radio observations, which would trace the low-energy tail of the synchrotron spectrum from the relic population, could further constrain the relative contributions of the two zones, as has been explored for other PWNe \citep{HAWCJ1809}.

\section{Conclusions} \label{sec:conclusions}

In this analysis of the pulsar wind nebula HAWC J1849-0000, which utilized 2860 days of HAWC Observatory data and other multi-wavelength data, the source is confirmed as a bright and slightly extended source. Its spectrum, reaching beyond 100 TeV, marks it as one of the most powerful UHE gamma-ray sources in the galaxy.

By performing time-dependent leptonic modeling of the broadband spectral energy distribution, we have successfully constrained the key physical parameters of the system. The PWN is a PeV electron accelerator candidate, capable of accelerating leptons to a maximum energy of $E_{\mathrm{cut}} \geq 2.9~\mathrm{PeV}$. Adopting an age of $\tau=9~\mathrm{kyr}$, the model yields a magnetic field of $B\left(\tau \right) \approx 4.28 ~\mathrm{\mu G}$ and an efficiency of $\eta\approx 0.63$ for converting the pulsar's spin-down power into relativistic electrons, indicating that the majority of the pulsar's rotational energy is channeled into the electron population responsible for the observed non-thermal emissions. 

Future observations with next-generation instruments will be essential to further resolve the nebula's structure and probe the detailed physics of the extreme acceleration.

\begin{acknowledgments} 

We thank Chanho Kim and Jaegeun Park for providing the X-ray flux points. We acknowledge the support from: the US National Science Foundation (NSF); the US Department of Energy Office of High-Energy Physics; the Laboratory Directed Research and Development (LDRD) program of Los Alamos National Laboratory; Consejo Nacional de Ciencia y Tecnolog\'{i}a (CONACyT), M\'{e}xico, grants LNC-2023-117, 271051, 232656, 260378, 179588, 254964, 258865, 243290, 132197, A1-S-46288, A1-S-22784, CF-2023-I-645, CBF2023-2024-1630, c\'{a}tedras 873, 1563, 341, 323, Red HAWC, M\'{e}xico; DGAPA-UNAM grants IG101323, IN111716-3, IN111419, IA102019, IN106521, IN114924, IN110521 , IN102223; VIEP-BUAP; PIFI 2012, 2013, PROFOCIE 2014, 2015; the University of Wisconsin Alumni Research Foundation; the Institute of Geophysics, Planetary Physics, and Signatures at Los Alamos National Laboratory; Polish Science Centre grant, 2024/53/B/ST9/02671; Coordinaci\'{o}n de la Investigaci\'{o}n Cient\'{i}fica de la Universidad Michoacana; Royal Society - Newton Advanced Fellowship 180385; Gobierno de España and European Union-NextGenerationEU, grant CNS2023- 144099; The Program Management Unit for Human Resources \& Institutional Development, Research and Innovation, NXPO (grant number B16F630069); Coordinaci\'{o}n General Acad\'{e}mica e Innovaci\'{o}n (CGAI-UdeG), PRODEP-SEP UDG-CA-499; Institute of Cosmic Ray Research (ICRR), University of Tokyo. H.F. acknowledges support by NASA under award number 80GSFC21M0002. C.R. acknowledges support from National Research Foundation of Korea (RS-2023-00280210, RS-2026-25470489). Y.S. acknowledges support from National Research Foundation of Korea (2018R1A6A1A06024977, RS-2021-NR058944, RS-2023-NR076954). We also acknowledge the significant contributions over many years of Stefan Westerhoff, Gaurang Yodh and Arnulfo Zepeda Dom\'inguez, all deceased members of the HAWC collaboration. Thanks to Scott Delay, Luciano D\'{i}az and Eduardo Murrieta for technical support.
\end{acknowledgments}

\begin{contribution}
Y.~Son performed the maximum likelihood analysis, GAMERA analysis, and prepared the original manuscript. R.~Babu wrote the GAMERA code about time-dependent leptonic model powered by the pulsar's rotational energy. C.D.~Rho performed data analysis and finalized the submitted manuscript. The full HAWC collaboration has contributed through the construction, calibration, and operation of the detector; the development and maintenance of reconstruction and analysis software; and vetting of the analysis presented in this manuscript. All authors have reviewed, discussed, and commented on the results and the manuscript.

\end{contribution}

\appendix
\restartappendixnumbering

\section{Morphological Models} \label{sec:specmorph}

This appendix provides the mathematical formulae for the morphological models used in our analysis.

\paragraph{2D Gaussian Distribution}
The 2D Gaussian morphology is defined by:
\begin{equation}
    f(\mathbf{x}) = \frac{1}{2\pi \sqrt{\det \Sigma}} \exp \left(-0.5(\mathbf{x}-\mathbf{x_{0}})^{\intercal} \Sigma^{-1} (\mathbf{x}-\mathbf{x_{0}})\right),
\end{equation}
where $\mathbf{x}_{0}$ is the vector representing the source center (RA, Dec), and $\Sigma$ is the covariance matrix. The covariance matrix is parameterized by the major axis length ($\sigma$), eccentricity ($e$), and position angle ($\theta$) as $\Sigma = U\Lambda U^{\intercal}$, where:
\begin{equation}
    \Lambda = \left( \begin{array}{cc} \sigma^2 & 0 \\ 0 & \sigma^2 (1-e^2) \end{array}\right) ~\mathrm{and~}~
         \\ U = \left( \begin{array}{cc} \cos \theta & -\sin \theta \\ \sin \theta & cos \theta \end{array}\right),
\end{equation}

For a radially symmetric Gaussian, the eccentricity $e$ is fixed to 0. For an elliptical Gaussian, $e$ and $\theta$ are treated as free parameters \citep{astromodels}.

\paragraph{2D Laplace Distribution}
The 2D Laplace distribution, which has a sharper core and wider tails than a Gaussian, was also tested \citep{gammapy}. Its morphology is defined as:
\begin{equation}
    f(\mathbf{r}) = \frac{1}{2\pi \sigma^{2}} \exp \left(-\frac{|\mathbf{x}-\mathbf{x_{0}}|}{\sigma}\right).
\end{equation}

\bibliography{sample7}{}

\begin{thebibliography}{}
\expandafter\ifx\csname natexlab\endcsname\relax\def\natexlab#1{#1}\fi
\providecommand{\url}[1]{\href{#1}{#1}}
\providecommand{\dodoi}[1]{doi:~\href{http://doi.org/#1}{\nolinkurl{#1}}}
\providecommand{\doeprint}[1]{\href{http://ascl.net/#1}{\nolinkurl{http://ascl.net/#1}}}
\providecommand{\doarXiv}[1]{\href{https://arxiv.org/abs/#1}{\nolinkurl{https://arxiv.org/abs/#1}}}

\bibitem[{A.~U. Abeysekara {et~al.}(2021)Abeysekara, Albert, Alfaro, {et~al.}}]{HAL}
Abeysekara, A.~U., Albert, A., Alfaro, R., {et~al.} 2021, in ICRC2021, 828, \dodoi{10.22323/1.395.0828}

\bibitem[{A.~U. Abeysekara {et~al.}(2019)Abeysekara, Albert, Alfaro, Alvarez, Álvarez, Camacho, Arceo, Arteaga-Velázquez, Arunbabu, Rojas, Solares, Baghmanyan, Belmont-Moreno, BenZvi, Brisbois, Caballero-Mora, Capistrán, Carramiñana, Casanova, Cotti, Cotzomi, de~León, Fuente, León, Dichiara, Dingus, DuVernois, Díaz-Vélez, Ellsworth, Engel, Espinoza, Fick, Fleischhack, Fraija, Galván-Gámez, García-González, Garfias, González, Goodman, Harding, Hernandez, Hinton, Hona, Hueyotl-Zahuantitla, Hui, Hüntemeyer, Iriarte, Jardin-Blicq, Joshi, Kaufmann, Kieda, Lara, Lee, Vargas, Linnemann, Longinotti, Luis-Raya, Lundeen, Malone, Marinelli, Martinez, Martinez-Castellanos, Martínez-Castro, Martínez-Huerta, Matthews, Miranda-Romagnoli, Morales-Soto, Moreno, Mostafá, Nayerhoda, Nellen, Newbold, Nisa, Noriega-Papaqui, Peisker, Pérez-Pérez, Pretz, Ren, Rho, Rivière, Rosa-González, Rosenberg, Ruiz-Velasco, Salazar, Greus, Sandoval, Schneider, Schoorlemmer, Arroyo, Sinnis, Smith, Springer, Surajbali,
  Tabachnick, Tanner, Tibolla, Tollefson, Torres, Weisgarber, Westerhoff, Wood, Yapici, Zepeda, \& Zhou}]{HAWCEE}
Abeysekara, A.~U., Albert, A., Alfaro, R., {et~al.} 2019, \bibinfo{title}{Measurement of the Crab Nebula Spectrum Past 100 TeV with HAWC,} The Astrophysical Journal, 881, 134, \dodoi{10.3847/1538-4357/ab2f7d}

\bibitem[{A.~U. Abeysekara {et~al.}(2020)Abeysekara, Albert, Alfaro, Angeles~Camacho, Arteaga-Vel\'azquez, Arunbabu, Avila~Rojas, Ayala~Solares, Baghmanyan, Belmont-Moreno, BenZvi, Brisbois, Caballero-Mora, Capistr\'an, Carrami\~nana, Casanova, Cotti, Cotzomi, Couti\~no~de Le\'on, De~la Fuente, de~Le\'on, Dichiara, Dingus, DuVernois, D\'{\i}az-V\'elez, Ellsworth, Engel, Espinoza, Fleischhack, Fraija, Galv\'an-G\'amez, Garcia, Garc\'{\i}a-Gonz\'alez, Garfias, Gonz\'alez, Goodman, Harding, Hernandez, Hinton, Hona, Huang, Hueyotl-Zahuantitla, H\"untemeyer, Iriarte, Jardin-Blicq, Joshi, Kaufmann, Kieda, Lara, Lee, Le\'on~Vargas, Linnemann, Longinotti, Luis-Raya, Lundeen, L\'opez-Coto, Malone, Marinelli, Martinez, Martinez-Castellanos, Mart\'{\i}nez-Castro, Mart\'{\i}nez-Huerta, Matthews, Miranda-Romagnoli, Morales-Soto, Moreno, Mostaf\'a, Nayerhoda, Nellen, Newbold, Nisa, Noriega-Papaqui, Peisker, P\'erez-P\'erez, Pretz, Ren, Rho, Rivi\`ere, Rosa-Gonz\'alez, Rosenberg, Ruiz-Velasco, Salesa~Greus, Sandoval,
  Schneider, Schoorlemmer, Sinnis, Smith, Springer, Surajbali, Tabachnick, Tanner, Tibolla, Tollefson, Torres, Torres-Escobedo, Villase\~nor, Weisgarber, Wood, Yapici, Zhang, \& Zhou}]{eHWC}
Abeysekara, A.~U., Albert, A., Alfaro, R., {et~al.} 2020, \bibinfo{title}{Multiple Galactic Sources with Emission Above 56 TeV Detected by HAWC,} Phys. Rev. Lett., 124, 021102, \dodoi{10.1103/PhysRevLett.124.021102}

\bibitem[{A.~U. {Abeysekara} {et~al.}(2023){Abeysekara}, {Albert}, {Alfaro}, {Alvarez}, {{\'A}lvarez}, {Araya}, {Arteaga-Vel{\'a}zquez}, {Arunbabu}, {Avila Rojas}, {Ayala Solares}, {Babu}, {Barber}, {Becerril}, {Belmont-Moreno}, {BenZvi}, {Blanco}, {Braun}, {Brisbois}, {Caballero-Mora}, {Cabrera Mart{\'\i}nez}, {Capistr{\'a}n}, {Carrami{\~n}ana}, {Casanova}, {Castillo}, {Chaparro-Amaro}, {Cotti}, {Cotzomi}, {Couti{\~n}o de Le{\'o}n}, {de la Fuente}, {de Le{\'o}n}, {De Young}, {Hernandez}, {Dingus}, {DuVernois}, {Durocher}, {D{\'\i}az-V{\'e}lez}, {Ellsworth}, {Engel}, {Espinoza}, {Fan}, {Fang}, {Fick}, {Fleischhack}, {Flores}, {Fraija}, {Garc{\'\i}a-Gonz{\'a}lez}, {Garcia-Torales}, {Garfias}, {Giacinti}, {Goksu}, {Gonz{\'a}lez}, {Gonz{\'a}lez-Mu{\~n}oz}, {Goodman}, {Harding}, {Hernandez}, {Hernandez}, {Hinton}, {Hona}, {Huang}, {Hueyotl-Zahuantitla}, {Hui}, {Humensky}, {H{\"u}ntemeyer}, {Iriarte}, {Imran}, {Jardin-Blicq}, {Joshi}, {Kaufmann}, {Kieda}, {Kunde}, {Lara}, {Lauer}, {Lee}, {Lennarz}, {Vargas},
  {Linnemann}, {Longinotti}, {Luis-Raya}, {Lundeen}, {Malone}, {Marandon}, {Marinelli}, {Martinez}, {Mart{\'\i}nez-Castellanos}, {Mart{\'\i}nez-Castro}, {Mart{\'\i}nez-Huerta}, {Matthews}, {Miranda-Romagnoli}, {Montaruli}, {Morales-Soto}, {Moreno}, {Mostaf{\'a}}, {Nayerhoda}, {Nellen}, {Newbold}, {Nisa}, {Noriega-Papaqui}, {Oceguera-Becerra}, {Olivera-Nieto}, {Omodei}, {Peisker}, {P{\'e}rez Araujo}, {P{\'e}rez-P{\'e}rez}, {Ponce}, {Pretz}, {Rho}, {Rosa-Gonz{\'a}lez}, {Ruiz-Velasco}, {Salazar}, {Salazar-Gallegos}, {Salesa Greus}, {Sandoval}, {Schneider}, {Schoorlemmer}, {Serna-Franco}, {Sinnis}, {Smith}, {Son}, {Sparks Woodle}, {Springer}, {Taboada}, {Tepe}, {Tibolla}, {Tollefson}, {Torres}, {Torres-Escobedo}, {Turner}, {Ure{\~n}a-Mena}, {Ukwatta}, {Varela}, {Vargas-Maga{\~n}a}, {Villase{\~n}or}, {Wang}, {Watson}, {Werner}, {Westerhoff}, {Willox}, {Wisher}, {Wood}, {Yodh}, {Zaborov}, {Zepeda}, {Zhou}, \& {HAWC Collaboration}}]{HAWCNIMA}
{Abeysekara}, A.~U., {Albert}, A., {Alfaro}, R., {et~al.} 2023, \bibinfo{title}{{The High-Altitude Water Cherenkov (HAWC) observatory in M{\'e}xico: The primary detector},} Nuclear Instruments and Methods in Physics Research A, 1052, 168253, \dodoi{10.1016/j.nima.2023.168253}

\bibitem[{A. Acharyya {et~al.}(2025)Acharyya, Adelfio, Ajello, Baldini, Ballet, Bartolini, Becerra~Gonzalez, Bellazzini, Bissaldi, Bonino, Bruel, Cameron, Caraveo, Casaburo, Casini, Castro, Cavazzuti, Ciprini, Cozzolongo, Cristarella~Orestano, Cuna, Cutini, D’Ammando, Depalo, Di~Lalla, Dinesh, Di~Venere, Domínguez, Eagle, Fiori, Fukazawa, Funk, Fusco, Gargano, Gasbarra, Gasparrini, Germani, Giacchino, Giglietto, Giliberti, Giordano, Giroletti, Green, Grenier, Grondin, Guiriec, Gupta, Harding, Hashizume, Hays, Hewitt, Horan, Hou, Kayanoki, Kuss, Laviron, Lemoine-Goumard, Liguori, Li, Liodakis, Loizzo, Longo, Loparco, Lorusso, Lovellette, Lubrano, Maldera, Malyshev, Martí-Devesa, Mazziotta, Mereu, Michelson, Mirabal, Mizuno, Monti-Guarnieri, Monzani, Morselli, Moskalenko, Omodei, Orlando, Paneque, Panzarini, Persic, Pesce-Rollins, Pillera, Porter, Principe, Rainò, Rando, Razzano, Reimer, Reimer, Sánchez-Conde, Saz~Parkinson, Serini, Sgrò, Siskind, Spandre, Spinelli, Strong, Tajima, Thayer, Tibaldo,
  Torres, Valverde, Wood, Zaharijas, \& Zhang}]{FermiPWN}
Acharyya, A., Adelfio, A., Ajello, M., {et~al.} 2025, \bibinfo{title}{A Systematic Search for MeV–GeV Pulsar Wind Nebulae without Gamma-Ray Detected Pulsars,} The Astrophysical Journal, 989, 110, \dodoi{10.3847/1538-4357/ade8f0}

\bibitem[{A. Albert {et~al.}(2020)Albert {et~al.}}]{3HWC}
Albert, A., {et~al.} 2020, \bibinfo{title}{{3HWC: The Third HAWC Catalog of Very-High-Energy Gamma-ray Sources},} Astrophys. J., 905, 76, \dodoi{10.3847/1538-4357/abc2d8}

\bibitem[{A. Albert {et~al.}(2021)Albert {et~al.}}]{HAWCJ2019}
Albert, A., {et~al.} 2021, \bibinfo{title}{{Spectrum and Morphology of the Very-high-energy Source HAWC J2019+368},} Astrophys. J., 911, 143, \dodoi{10.3847/1538-4357/abecda}

\bibitem[{A. {Albert} {et~al.}(2024){Albert}, {Alfaro}, {Alvarez}, {Andres}, {Arteaga Velazquez}, {Avila Rojas}, {Ayala Solares}, {Babu}, {Belmont-Moreno}, {Capistr{\'a}n Rojas}, {Yun}, {Carrami{\~n}ana}, {Carreon-Gonzalez}, {Cotti}, {Cotzomi}, {Couti{\~n}o de Le{\'o}n}, {de la Fuente}, {Depaoli}, {de Le{\'o}n}, {Diaz Hernandez}, {D{\'\i}az V{\'e}lez}, {Dingus}, {Durocher}, {DuVernois}, {Engel}, {Espinoza Hern{\'a}ndez}, {Fan}, {Fang}, {Fraija}, {Garcia-Gonzalez}, {Garfias}, {Goksu}, {Gonz{\'a}lez}, {Goodman}, {Groetsch}, {Harding}, {Hern{\'a}ndez Cadena}, {Herzog}, {Hinton}, {Hona}, {Huang}, {Hueyotl-Zahuantitla}, {H{\"u}ntemeyer}, {Iriarte}, {Joshi}, {Kaufmann}, {Kieda}, {Lara}, {Lee}, {Lee}, {Vargas}, {Linnemann}, {Longinotti}, {Luis-Raya}, {Malone}, {Mart{\'\i}nez-Castro}, {Matthews}, {Miranda-Romagnoli}, {Montes}, {Morales Soto}, {Mostafa}, {Nellen}, {Nisa}, {Noriega-Papaqui}, {Olivera-Nieto}, {Omodei}, {P{\'e}rez Araujo}, {P{\'e}rez P{\'e}rez}, {Pratts}, {Rho}, {Rosa-Gonzalez}, {Ruiz-Velasco},
  {Salazar}, {Salazar-Gallegos}, {Sandoval}, {Schneider}, {Schwefer}, {Serna-Franco}, {Smith}, {Son}, {Springer}, {Tibolla}, {Tollefson}, {Torres}, {Torres Escobedo}, {Turner}, {Ure{\~n}a-Mena}, {Varela}, {Villase{\~n}or}, {Wang}, {Watson}, {Werner}, {Whitaker}, {Willox}, {Hongyi Wu}, {Zhou}, \& {Caballero Mora}}]{2eHWC}
{Albert}, A., {Alfaro}, R., {Alvarez}, C., {et~al.} 2024, in ICRC2023, 698

\bibitem[{A. Albert {et~al.}(2024{\natexlab{a}})Albert, Alfaro, Alvarez, Andrés, Arteaga-Velázquez, Avila~Rojas, Ayala~Solares, Babu, Belmont-Moreno, Bernal, Caballero-Mora, Capistrán, Carramiñana, Carreón, Casanova, Cotti, Cotzomi, Coutiño~de León, De~la Fuente, de~León, Depaoli, Di~Lalla, Díaz~Hernández, Dingus, DuVernois, Engel, Ergin, Espinoza, Fan, Fang, Fraija, Fraija, García-González, Garfias, Goksu, González, Goodman, Groetsch, Harding, Hernández-Cadena, Herzog, Hinton, Huang, Hueyotl-Zahuantitla, Hüntemeyer, Iriarte, Kaufmann, Lara, Lee, León~Vargas, Linnemann, Longinotti, Luis-Raya, Malone, Martínez-Castro, Matthews, Miranda-Romagnoli, Montes, Moreno, Mostafá, Nellen, Nisa, Noriega-Papaqui, Olivera-Nieto, Omodei, Osorio-Archila, Pérez~Araujo, Pérez-Pérez, Rho, Rosa-González, Ruiz-Velasco, Salazar, Salazar-Gallegos, Sandoval, Schneider, Schwefer, Serna-Franco, Smith, Son, Springer, Tibolla, Tollefson, Torres, Torres-Escobedo, Turner, Ureña-Mena, Varela, Wang, Watson, Whitaker,
  Willox, Wu, Yu, Yun-Cárcamo, Zhou, \& Collaboration}]{HAWCPass5}
Albert, A., Alfaro, R., Alvarez, C., {et~al.} 2024{\natexlab{a}}, \bibinfo{title}{Performance of the HAWC Observatory and TeV Gamma-Ray Measurements of the Crab Nebula with Improved Extensive Air Shower Reconstruction Algorithms,} \apj, 972, 144, \dodoi{10.3847/1538-4357/ad5f2d}

\bibitem[{A. Albert {et~al.}(2024{\natexlab{b}})Albert, Alfaro, Alvarez, Arteaga-Velázquez, Avila~Rojas, Babu, Belmont-Moreno, Bernal, Breuhaus, Caballero-Mora, Capistrán, Carramiñana, Casanova, Cotzomi, De~la Fuente, Depaoli, Di~Lalla, Diaz~Hernandez, Dingus, DuVernois, Espinoza, Fan, Fang, Fick, Fraija, García-González, Garfias, Gonzalez~Muñoz, González, Goodman, Groetsch, Harding, Hernández-Cadena, Herzog, Huang, Hueyotl-Zahuantitla, Hüntemeyer, Iriarte, Joshi, Kaufmann, Lara, Lee, León~Vargas, Longinotti, Luis-Raya, Malone, Martínez-Castro, Matthews, Miranda-Romagnoli, Montes, Morales-Soto, Moreno, Mostafá, Nellen, Newbold, Nisa, Noriega-Papaqui, Osorio, Pérez~Araujo, Pérez-Pérez, Rho, Rosa-González, Ruiz-Velasco, Salazar, Sandoval, Schneider, Serna-Franco, Smith, Son, Springer, Tibolla, Tollefson, Torres, Torres-Escobedo, Turner, Ureña-Mena, Varela, Wang, Watson, Willox, Yun-Cárcamo, Zhou, \& COLLABORATION)}]{HAWCJ1809}
Albert, A., Alfaro, R., Alvarez, C., {et~al.} 2024{\natexlab{b}}, \bibinfo{title}{TeV Analysis of a Source-rich Region with the HAWC Observatory: Is HESS J1809-193 a Potential Hadronic PeVatron?} The Astrophysical Journal, 972, 21, \dodoi{10.3847/1538-4357/ad59a6}

\bibitem[{R. Alfaro {et~al.}(2024)Alfaro, Alvarez, Arteaga-Velázquez, Avila~Rojas, Ayala~Solares, Babu, Belmont-Moreno, Bernal, Caballero-Mora, Capistrán, Carramiñana, Casanova, Cotti, Cotzomi, Coutiño~de León, De~la Fuente, de~León, Depaoli, Desiati, Di~Lalla, Diaz~Hernandez, Dingus, DuVernois, Engel, Ergin, Espinoza, Fan, Fang, Fraija, Fraija, García-González, Garfias, González, Goodman, Groetsch, Harding, Hernández-Cadena, Herzog, Hinton, Huang, Hueyotl-Zahuantitla, Humensky, Hüntemeyer, Kaufmann, Kieda, Lee, Lee, León~Vargas, Linnemann, Longinotti, Luis-Raya, Malone, Martinez, Martínez-Castro, Matthews, Miranda-Romagnoli, Montes, Moreno, Mostafá, Najafi, Nellen, Nisa, Olivera-Nieto, Omodei, Pérez~Araujo, Pérez-Pérez, Rho, Rosa-González, Salazar, Salazar-Gallegos, Sandoval, Schneider, Serna-Franco, Smith, Son, Springer, Tibolla, Tollefson, Torres, Torres-Escobedo, Turner, Ureña-Mena, Varela, Villaseñor, Wang, Wang, Watson, Willox, Yu, Yun-Cárcamo, \& Zhou}]{Boomerang}
Alfaro, R., Alvarez, C., Arteaga-Velázquez, J.~C., {et~al.} 2024, \bibinfo{title}{Testing the molecular cloud paradigm for ultra-high-energy gamma ray emission from the direction of SNR G106.3+2.7,} \aap, 691, A89, \dodoi{10.1051/0004-6361/202451514}

\bibitem[{M. Amenomori {et~al.}(2023)Amenomori, Asano, Bao, Bi, Chen, Chen, Chen, Chen, Chen, Cirennima, Cui, Danzengluobu, Ding, Fang, Fang, Feng, Feng, Feng, Gao, Gomi, Gou, Guo, Guo, Hayashi, He, He, Hibino, Hotta, Hu, Hu, Hu, Huang, Jia, Jiang, Jiang, Jin, Kasahara, Katayose, Kato, Kato, Kawahara, Kawashima, Kawata, Kozai, Kurashige, Labaciren, Le, Li, Li, Li, Li, Lin, Liu, Liu, Liu, Liu, Liu, Liu, Lu, Meng, Meng, Munakata, Nagaya, Nakamura, Nakazawa, Nanjo, Ning, Nishizawa, Noguchi, Ohnishi, Okukawa, Ozawa, Qian, Qian, Qu, Saito, Sakakibara, Sakata, Sako, Sako, Sasaki, Shao, Shibata, Shiomi, Sugimoto, Takano, Takita, Tan, Tateyama, Torii, Tsuchiya, Udo, Wang, Wang, Wang, Wangdui, Wu, Wu, Xu, Xue, Yang, Yao, Yin, Yokoe, Yu, Yuan, Zhai, Zhang, Zhang, Zhang, Zhang, Zhang, Zhang, Zhang, Zhao, Zhaxisangzhu, Zhou, \& Zou}]{TibetPWN}
Amenomori, M., Asano, S., Bao, Y.~W., {et~al.} 2023, \bibinfo{title}{Observation of Gamma Rays up to 320 TeV from the Middle-aged TeV Pulsar Wind Nebula HESS J1849−000,} The Astrophysical Journal, 954, 200, \dodoi{10.3847/1538-4357/acebce}

\bibitem[{M. Araya \& J.~A. Álvarez Quesada(2024)Araya \& Álvarez Quesada}]{Araya2024}
Araya, M., \& Álvarez Quesada, J.~A. 2024, \bibinfo{title}{Discovery of an extended GeV counterpart to the TeV source 1LHAASO J1945+2424 in Fermi-LAT data,} Monthly Notices of the Royal Astronomical Society, 527, 8006, \dodoi{10.1093/mnras/stad3739}

\bibitem[{N. Ben~Bekhti {et~al.}(2016)Ben~Bekhti, Flöer, Keller, Kerp, Lenz, Winkel, Bailin, Calabretta, Dedes, Ford, Gibson, Haud, Janowiecki, Kalberla, Lockman, McClure-Griffiths, Murphy, Nakanishi, Pisano, \& Staveley-Smith}]{HI4PI}
Ben~Bekhti, N., Flöer, L., Keller, R., {et~al.} 2016, \bibinfo{title}{HI4PI: a full-sky HI survey based on EBHIS and GASS,} \aap, 594, A116, \dodoi{10.1051/0004-6361/201629178}

\bibitem[{Z. Cao {et~al.}(2021)Cao {et~al.}}]{UHELHAASO}
Cao, Z., {et~al.} 2021, \bibinfo{title}{{Ultrahigh-energy photons up to 1.4 petaelectronvolts from 12 $\gamma$-ray Galactic sources},} Nature, 594, 33, \dodoi{10.1038/s41586-021-03498-z}

\bibitem[{Z. {Cao} {et~al.}(2021){Cao}, {Aharonian}, {An}, {Axikegu}, {Bai}, {Bai}, {Bao}, {Bastieri}, {Bi}, {Bi}, {Cai}, {Cai}, {Cao}, {Chang}, {Chang}, {Chen}, {Chen}, {Chen}, {Chen}, {Chen}, {Chen}, {Chen}, {Chen}, {Chen}, {Chen}, {Chen}, {Chen}, {Chen}, {Chen}, {Cheng}, {Cheng}, {Cui}, {Cui}, {Cui}, {D'Ettorre Piazzoli}, {Dai}, {Dai}, {Dai}, {Danzengluobu}, {Della Volpe}, {Dong}, {Duan}, {Fan}, {Fan}, {Fan}, {Fang}, {Fang}, {Feng}, {Feng}, {Feng}, {Feng}, {Gao}, {Gao}, {Gao}, {Gao}, {Gao}, {Ge}, {Geng}, {Gong}, {Gou}, {Gu}, {Guo}, {Guo}, {Guo}, {Guo}, {Guo}, {Han}, {He}, {He}, {He}, {He}, {He}, {He}, {Heller}, {Hor}, {Hou}, {Hou}, {Hu}, {Hu}, {Hu}, {Hu}, {Huang}, {Huang}, {Huang}, {Huang}, {Huang}, {Huang}, {Ji}, {Ji}, {Jia}, {Jiang}, {Jiang}, {Jin}, {Ke}, {Kuleshov}, {Levochkin}, {Li}, {Li}, {Li}, {Li}, {Li}, {Li}, {Li}, {Li}, {Li}, {Li}, {Li}, {Li}, {Li}, {Li}, {Li}, {Li}, {Li}, {Li}, {Liang}, {Liang}, {Lin}, {Liu}, {Liu}, {Liu}, {Liu}, {Liu}, {Liu}, {Liu}, {Liu}, {Liu}, {Liu}, {Liu}, {Liu}, {Liu},
  {Liu}, {Liu}, {Liu}, {Long}, {Lu}, {Lv}, {Ma}, {Ma}, {Ma}, {Mao}, {Masood}, {Min}, {Mitthumsiri}, {Montaruli}, {Nan}, {Pang}, {Pattarakijwanich}, {Pei}, {Qi}, {Qi}, {Qiao}, {Qin}, {Ruffolo}, {Rulev}, {Saiz}, {Shao}, {Shchegolev}, {Sheng}, {Shi}, {Song}, {Stenkin}, {Stepanov}, {Su}, {Sun}, {Sun}, {Sun}, {Tam}, {Tang}, {Tian}, {Wang}, {Wang}, {Wang}, {Wang}, {Wang}, {Wang}, {Wang}, {Wang}, {Wang}, {Wang}, {Wang}, {Wang}, {Wang}, {Wang}, {Wang}, {Wang}, {Wang}, {Wang}, {Wang}, {Wang}, {Wang}, {Wang}, {Wei}, {Wei}, {Wei}, {Wen}, {Wu}, {Wu}, {Wu}, \& {Wu}}]{UHELHAASOCrab}
{Cao}, Z., {Aharonian}, F., {An}, Q., {et~al.} 2021, \bibinfo{title}{{Peta-electron volt gamma-ray emission from the Crab Nebula},} Science, 373, 425, \dodoi{10.1126/science.abg5137}

\bibitem[{Z. {Cao} {et~al.}(2023){Cao}, {Aharonian}, {An}, {Axikegu}, {Bai}, {Bao}, {Bastieri}, {Bi}, {Bi}, {Cai}, {Cao}, {Cao}, {Cao}, {Chang}, {Chang}, {Chen}, {Chen}, {Chen}, {Chen}, {Chen}, {Chen}, {Chen}, {Chen}, {Chen}, {Chen}, {Chen}, {Chen}, {Cheng}, {Cheng}, {Cui}, {Cui}, {Cui}, {Cui}, {Dai}, {Dai}, {Dai}, {Danzengluobu}, {della Volpe}, {Dong}, {Duan}, {Fan}, {Fan}, {Fang}, {Fang}, {Feng}, {Feng}, {Feng}, {Feng}, {Feng}, {Gabici}, {Gao}, {Gao}, {Gao}, {Gao}, {Gao}, {Gao}, {Ge}, {Geng}, {Giacinti}, {Gong}, {Gou}, {Gu}, {Guo}, {Guo}, {Guo}, {Guo}, {Han}, {He}, {He}, {He}, {He}, {He}, {Heller}, {Hor}, {Hou}, {Hou}, {Hou}, {Hu}, {Hu}, {Hu}, {Huang}, {Huang}, {Huang}, {Huang}, {Huang}, {Huang}, {Huang}, {Ji}, {Jia}, {Jia}, {Jiang}, {Jiang}, {Jiang}, {Jin}, {Kang}, {Ke}, {Kuleshov}, {Kurinov}, {Li}, {Li}, {Li}, {Li}, {Li}, {Li}, {Li}, {Li}, {Li}, {Li}, {Li}, {Li}, {Li}, {Li}, {Li}, {Li}, {Li}, {Li}, {Li}, {Liang}, {Liang}, {Lin}, {Liu}, {Liu}, {Liu}, {Liu}, {Liu}, {Liu}, {Liu}, {Liu}, {Liu}, {Liu}, {Liu},
  {Liu}, {Liu}, {Liu}, {Lu}, {Luo}, {Lv}, {Ma}, {Ma}, {Ma}, {Mao}, {Min}, {Mitthumsiri}, {Mu}, {Nan}, {Neronov}, {Ou}, {Pang}, {Pattarakijwanich}, {Pei}, {Qi}, {Qi}, {Qiao}, {Qin}, {Ruffolo}, {S{\'a}iz}, {Semikoz}, {Shao}, {Shao}, {Shchegolev}, {Sheng}, {Shu}, {Song}, {Stenkin}, {Stepanov}, {Su}, {Sun}, {Sun}, {Sun}, {Tam}, {Tang}, {Tang}, {Tian}, {Wang}, {Wang}, {Wang}, {Wang}, {Wang}, {Wang}, {Wang}, {Wang}, {Wang}, {Wang}, {Wang}, {Wang}, {Wang}, {Wang}, {Wang}, {Wang}, {Wang}, {Wang}, {Wang}, {Wang}, {Wang}, {Wei}, {Wei}, {Wei}, {Wen}, {Wu}, {Wu}, {Wu}, {Wu}, {Wu}, {Xi}, {Xia}, {Xia}, {Xiang}, {Xiao}, {Xiao}, {Xin}, {Xin}, {Xing}, {Xiong}, {Xu}, {Xu}, {Xu}, {Xu}, {Xue}, {Yan}, {Yan}, {Yan}, {Yang}, {Yang}, {Yang}, {Yang}, {Yang}, {Yang}, {Yang}, {Yang}, {Yang}, {Yao}, {Yao}, {Ye}, {Yin}, {Yin}, {You}, {You}, {Yu}, {Yuan}, {Yue}, {Zeng}, {Zeng}, {Zeng}, {Zha}, {Zhang}, {Zhang}, {Zhang}, {Zhang}, {Zhang}, {Zhang}, {Zhang}, {Zhang}, {Zhang}, {Zhang}, {Zhang}, {Zhang}, {Zhang}, {Zhang}, {Zhang}, {Zhang},
  {Zhang}, {Zhang}, {Zhao}, {Zhao}, {Zhao}, {Zhao}, {Zhao}, {Zheng}, {Zhou}, {Zhou}, {Zhou}, {Zhou}, {Zhou}, {Zhou}, {Zhou}, {Zhu}, {Zhu}, {Zhu}, {Zhu}, \& {Zuo.}}]{1LHAASO}
{Cao}, Z., {Aharonian}, F., {An}, Q., {et~al.} 2023, \bibinfo{title}{{The First LHAASO Catalog of Gamma-Ray Sources},} arXiv e-prints, arXiv:2305.17030, \dodoi{10.48550/arXiv.2305.17030}

\bibitem[{Z. Cao {et~al.}(2025)Cao, Aharonian, Axikegu, Bai, Bao, Bastieri, Bi, Bi, Bian, Bukevich, Cao, Cao, Cao, Chang, Chang, Chen, Chen, Chen, Chen, Chen, Chen, Chen, Chen, Chen, Chen, Chen, Chen, Chen, Chen, Cheng, Cheng, Chu, Cui, Cui, Cui, Cui, Dai, Dai, Dai, Danzengluobu, Dong, Duan, Fan, Fan, Fang, Fang, Fang, Feng, Feng, Feng, Feng, Feng, Feng, Feng, Gabici, Gao, Gao, Gao, Gao, Gao, Ge, Ge, Geng, Giacinti, Gong, Gou, Gu, Guo, Guo, Guo, Guo, Guo, Han, Hannuksela, Hasan, He, He, He, He, Hor, Hou, Hou, Hou, Hu, Hu, Hu, Huang, Huang, Huang, Huang, Huang, Huang, Huang, Huang, Ji, Jia, Jia, Jiang, Jiang, Jiang, Jiang, Jin, Kang, Karpikov, Khangulyan, Kuleshov, Kurinov, Li, Li, Li, Li, Li, Li, Li, Li, Li, Li, Li, Li, Li, Li, Li, Li, Li, Li, Li, Liang, Liang, Lin, Liu, Liu, Liu, Liu, Liu, Liu, Liu, Liu, Liu, Liu, Liu, Liu, Liu, Liu, Luo, Luo, Lv, Ma, Ma, Ma, Mao, Min, Mitthumsiri, Mu, Nan, Neronov, Ng, Ou, Pattarakijwanich, Pei, Qi, Qi, Qiao, Qin, Raza, Ruffolo, Sáiz, Saeed, Semikoz, Shao, Shchegolev,
  Sheng, Shu, Song, Stenkin, Stepanov, Su, Sun, Sun, Sun, Sun, Takata, Tam, Tang, Tang, Tang, Tian, Wan, Wang, Wang, Wang, Wang, Wang, Wang, Wang, Wang, Wang, Wang, Wang, Wang, Wang, Wang, Wang, Wang, Wang, Wang, Wang, Wang, Wang, Wang, Wei, Wei, Wei, Wen, Wu, Wu, Wu, Wu, Wu, Wu, Xi, Xia, Xiang, Xiao, Xiao, Xin, Xing, Xiong, Xiong, Xu, Xu, Xu, Xu, Xue, Yan, Yan, Yan, Yang, Yang, Yang, Yang, Yang, Yang, Yang, Yang, Yao, Yao, Yin, Yin, You, You, Yu, Yuan, Yue, Zeng, Zeng, Zeng, Zha, Zhang, Zhang, Zhang, Zhang, Zhang, Zhang, Zhang, Zhang, Zhang, Zhang, Zhang, Zhang, Zhang, Zhang, Zhang, Zhang, Zhang, Zhang, Zhao, Zhao, Zhao, Zhao, Zhao, Zhao, Zheng, Zhong, Zhou, Zhou, Zhou, Zhou, Zhou, Zhou, Zhou, Zhou, Zhu, Zhu, Zhu, Zhu, Zhu, Zou, \& Zuo}]{LHAASOW43}
Cao, Z., Aharonian, F., Axikegu, {et~al.} 2025, \bibinfo{title}{Observation of the γ-ray emission from W43 with LHAASO,} Science China Physics, Mechanics \& Astronomy, 68, \dodoi{10.1007/s11433-024-2477-9}

\bibitem[{H.~E.~S.~S. Collaboration(2018)Collaboration}]{HGPS}
Collaboration, H.~E.~S.~S. 2018, \bibinfo{title}{{The H.E.S.S. Galactic plane survey},} \aap, 612, A1, \dodoi{10.1051/0004-6361/201732098}

\bibitem[{T.~M. {Dame} {et~al.}(2001){Dame}, {Hartmann}, \& {Thaddeus}}]{CfA}
{Dame}, T.~M., {Hartmann}, D., \& {Thaddeus}, P. 2001, \bibinfo{title}{{The Milky Way in Molecular Clouds: A New Complete CO Survey},} \apj, 547, 792, \dodoi{10.1086/318388}

\bibitem[{A. {Donath} {et~al.}(2023){Donath}, {Terrier}, {Remy}, {Sinha}, {Nigro}, {Pintore}, {Kh\'elifi}, {Olivera-Nieto}, {Ruiz}, {Br\"ugge}, {Linhoff}, {Contreras}, {Acero}, {Aguasca-Cabot}, {Berge}, {Bhattacharjee}, {Buchner}, {Boisson}, {Carreto Fidalgo}, {Chen}, {de Bony de Lavergne}, {de Miranda Cardoso}, {Deil}, {F\"u\ss{}ling}, {Funk}, {Giunti}, {Hinton}, {Jouvin}, {King}, {Lefaucheur}, {Lemoine-Goumard}, {Lenain}, {L\'opez-Coto}, {Mohrmann}, {Morcuende}, {Panny}, {Regeard}, {Saha}, {Siejkowski}, {Siemiginowska}, {Sip"ocz}, {Unbehaun}, {van Eldik}, {Vuillaume}, \& {Zanin}}]{gammapy}
{Donath}, A., {Terrier}, R., {Remy}, Q., {et~al.} 2023, \bibinfo{title}{Gammapy: A Python package for gamma-ray astronomy,} \aap, 678, A157, \dodoi{10.1051/0004-6361/202346488}

\bibitem[{A. Dundovic {et~al.}(2021)Dundovic, Evoli, Gaggero, \& Grasso}]{HERMES}
Dundovic, A., Evoli, C., Gaggero, D., \& Grasso, D. 2021, \bibinfo{title}{Simulating the Galactic multi-messenger emissions with HERMES,} \aap, 653, A18, \dodoi{10.1051/0004-6361/202140801}

\bibitem[{D. Foreman-Mackey {et~al.}(2013)Foreman-Mackey, Hogg, Lang, \& Goodman}]{emcee}
Foreman-Mackey, D., Hogg, D.~W., Lang, D., \& Goodman, J. 2013, \bibinfo{title}{emcee: The MCMC Hammer,} Publications of the Astronomical Society of the Pacific, 125, 306–312, \dodoi{10.1086/670067}

\bibitem[{B.~M. Gaensler \& P.~O. Slane(2006)Gaensler \& Slane}]{PWNEvolution}
Gaensler, B.~M., \& Slane, P.~O. 2006, \bibinfo{title}{The Evolution and Structure of Pulsar Wind Nebulae,} Annual Review of Astronomy and Astrophysics, 44, 17–47, \dodoi{10.1146/annurev.astro.44.051905.092528}

\bibitem[{E.~V. Gotthelf {et~al.}(2011)Gotthelf, Halpern, Terrier, \& Mattana}]{RXTE}
Gotthelf, E.~V., Halpern, J.~P., Terrier, R., \& Mattana, F. 2011, \bibinfo{title}{{Discovery of an Energetic 38.5 ms Pulsar Powering the Gamma-ray Source IGR J18490-0000/HESS J1849-000},} Astrophys. J. Lett., 729, L16, \dodoi{10.1088/2041-8205/729/2/L16}

\bibitem[{J. {Hahn} {et~al.}(2022){Hahn}, {Romoli}, \& {Breuhaus}}]{GAMERA}
{Hahn}, J., {Romoli}, C., \& {Breuhaus}, M. 2022, \bibinfo{title}{{GAMERA: Source modeling in gamma astronomy},} Astrophysics Source Code Library, record ascl:2203.007

\bibitem[{G. Hobbs {et~al.}(2004)Hobbs, Manchester, Teoh, \& Hobbs}]{ATNF}
Hobbs, G., Manchester, R., Teoh, A., \& Hobbs, M. 2004, \bibinfo{title}{{The atnf pulsar catalogue},} IAU Symp., 218, 139.
\newblock \doarXiv{astro-ph/0309219}

\bibitem[{S.~R. Kelner {et~al.}(2006)Kelner, Aharonian, \& Bugayov}]{Kelner}
Kelner, S.~R., Aharonian, F.~A., \& Bugayov, V.~V. 2006, \bibinfo{title}{Energy spectra of gamma rays, electrons, and neutrinos produced at proton-proton interactions in the very high energy regime,} Phys. Rev. D, 74, 034018, \dodoi{10.1103/PhysRevD.74.034018}

\bibitem[{C. Kim {et~al.}(2024)Kim, Park, Woo, Silverman, An, Bamba, Mori, Reynolds, \& Safi-Harb}]{XrayPWN}
Kim, C., Park, J., Woo, J., {et~al.} 2024, \bibinfo{title}{{X-Ray Characterization of the Pulsar PSR J1849\ensuremath{-}0001 and Its Wind Nebula G32.64+0.53 Associated with TeV Sources Detected by H.E.S.S., HAWC, Tibet AS\ensuremath{\gamma}, and LHAASO},} Astrophys. J., 960, 78, \dodoi{10.3847/1538-4357/ad0ecd}

\bibitem[{L. Nie {et~al.}(2022)Nie, Liu, Jiang, \& Geng}]{Nie_2022}
Nie, L., Liu, Y., Jiang, Z., \& Geng, X. 2022, \bibinfo{title}{Ultra-high-energy Gamma-Ray Radiation from the Crab Pulsar Wind Nebula,} The Astrophysical Journal, 924, 42, \dodoi{10.3847/1538-4357/ac348d}

\bibitem[{J. {Park} {et~al.}(2023){Park}, {Kim}, {Woo}, {An}, {Mori}, {Reynolds}, \& {Safi-Harb}}]{Park2023}
{Park}, J., {Kim}, C., {Woo}, J., {et~al.} 2023, \bibinfo{title}{{A Broadband X-Ray Study of the Rabbit Pulsar Wind Nebula Powered by PSR J1418-6058},} \apj, 945, 66, \dodoi{10.3847/1538-4357/acba0e}

\bibitem[{G. Schwarz(1978)Schwarz}]{BIC}
Schwarz, G. 1978, \bibinfo{title}{{Estimating the Dimension of a Model},} The Annals of Statistics, 6, 461 , \dodoi{10.1214/aos/1176344136}

\bibitem[{A. Vehtari {et~al.}(2021)Vehtari, Gelman, Simpson, Carpenter, \& B{\"u}rkner}]{RubinStatistic}
Vehtari, A., Gelman, A., Simpson, D., Carpenter, B., \& B{\"u}rkner, P.-C. 2021, \bibinfo{title}{{Rank-Normalization, Folding, and Localization: An Improved $\widehat{R}$ for Assessing Convergence of MCMC (with Discussion)},} Bayesian Analysis, 16, 667 , \dodoi{10.1214/20-BA1221}

\bibitem[{G. {Vianello} {et~al.}(2025){Vianello}, {Burgess}, {Fleischhack}, {Di Lalla}, \& {Omodei}}]{astromodels}
{Vianello}, G., {Burgess}, J.~M., {Fleischhack}, H., {Di Lalla}, N., \& {Omodei}, N. 2025, \bibinfo{title}{{astromodels: Spatial and spectral models for astrophysics},}, Astrophysics Source Code Library, record ascl:2506.019 \doeprint{2506.019}

\bibitem[{G. Vianello {et~al.}(2015)Vianello, Lauer, Younk, Tibaldo, Burgess, Ayala, Harding, Hui, Omodei, \& Zhou}]{threeML}
Vianello, G., Lauer, R.~J., Younk, P., {et~al.} 2015, \bibinfo{title}{The Multi-Mission Maximum Likelihood framework (3ML),} \doarXiv{1507.08343}

\bibitem[{S.~P. {Wakely} \& D. {Horan}(2008){Wakely} \& {Horan}}]{TeVCat}
{Wakely}, S.~P., \& {Horan}, D. 2008, in International Cosmic Ray Conference, Vol.~3, International Cosmic Ray Conference, 1341--1344

\bibitem[{S.~S. Wilks(1938)Wilks}]{Wilks}
Wilks, S.~S. 1938, \bibinfo{title}{{The Large-Sample Distribution of the Likelihood Ratio for Testing Composite Hypotheses},} The Annals of Mathematical Statistics, 9, 60 , \dodoi{10.1214/aoms/1177732360}

\bibitem[{B.-T. {Zhu} {et~al.}(2018){Zhu}, {Zhang}, \& {Fang}}]{PWNAge}
{Zhu}, B.-T., {Zhang}, L., \& {Fang}, J. 2018, \bibinfo{title}{{Multiband nonthermal radiative properties of pulsar wind nebulae},} \aap, 609, A110, \dodoi{10.1051/0004-6361/201629108}

\end{thebibliography}
\bibliographystyle{aasjournal}

\end{document}